%% file: draft.tex
\documentclass[aps,twocolumn,floatfix,superscriptaddress,longbibliography,nofootinbib]{revtex4}

\usepackage{graphicx}

\usepackage{multirow}
\usepackage{epsfig}
\usepackage{multirow}
\usepackage{comment}
\usepackage{hyperref}
\usepackage{transparent}

\usepackage{color}
\usepackage[usenames,dvipsnames]{xcolor}
\usepackage{amssymb}

\usepackage{amsfonts}
\usepackage{amsmath}

\usepackage{algorithm}
\usepackage{algpseudocode} 
\usepackage{setspace} 

\usepackage{ulem}

\usepackage[utf8]{inputenc}

\newcommand{\beq}{\begin{equation}}
\newcommand{\eeq}{\end{equation}}
\newcommand{\bea}{\begin{eqnarray}}
\newcommand{\eea}{\end{eqnarray}}
\newcommand{\ba}{\begin{array}}
\newcommand{\ea}{\end{array}}

\begin{document}

\title{Fast and Accurate Inference on Gravitational Waves from Precessing Compact Binaries}

\def\addCornell{Cornell Center for Astrophysics and Planetary Science, Cornell University, Ithaca, NY 14853, USA}

\def\addLIGO{LIGO, California Institute of Technology, Pasadena, CA 91125, USA}

\def\addAEI{Albert-Einstein-Institut, Max-Planck-Institut f\"{u}r Gravitationsphysik, D-14476 Golm, Germany}

\def\addBham{School of Physics and Astronomy, University of Birmingham, Edgbaston, Birmingham B15 2TT, United Kingdom}

\def\addTAPIR{TAPIR, Walter Burke Institute for Theoretical Physics, California Institute of Technology, Pasadena, CA 91125, USA}

\author{Rory Smith}
\email{rory.smith@caltech.edu}
\affiliation{\addLIGO}

\author{Scott E. Field}
\email{sfield@astro.cornell.edu}
\affiliation{\addCornell}

\author{Kent Blackburn}
\affiliation{\addLIGO}

\author{Carl-Johan Haster}
\affiliation{\addBham}

\author{Michael P\"{u}rrer}
\affiliation{\addAEI}

\author{Vivien Raymond}
\affiliation{\addAEI}

\author{Patricia Schmidt}
\affiliation{\addLIGO}
\affiliation{\addTAPIR}

\date\today

\begin{abstract}
Inferring astrophysical information from gravitational waves emitted by compact binaries is one of the key science goals of gravitational-wave astronomy.
In order to reach the full scientific potential of gravitational-wave experiments we require techniques to mitigate the cost of Bayesian inference, especially as gravitational-wave signal models and analyses become increasingly sophisticated and detailed.
Reduced order models (ROMs) of gravitational waveforms can significantly reduce the computational cost of inference by removing redundant computations. In this paper we construct the first reduced order models of gravitational-wave signals that include the effects of spin-precession, inspiral, merger, and ringdown in compact object binaries, and which are valid for component masses describing binary neutron star, binary black hole and mixed binary systems. This work utilizes the waveform model known as ``IMRPhenomPv2''. Our ROM enables the use of a fast \textit{reduced order quadrature} (ROQ) integration rule which allows us to approximate Bayesian probability density functions at a greatly reduced computational cost.
We find that the ROQ rule can be used to speed up inference by factors as high as 300 without introducing systematic bias.
This corresponds to a reduction in computational time from around half a year to a half a day, for the longest duration/lowest mass signals. The ROM and ROQ rule are available with the main inference library of the LIGO Scientific Collaboration, \texttt{LALInference}.
\end{abstract}


\maketitle

\section{Introduction}
\label{sec:intro}

With the first gravitational-wave (GW) detection reported in February 2016, an exciting era of GW astronomy has begun~\cite{PhysRevLett.116.061102}. The discovery of the GW source GW150914 with the advanced Laser Interferometer Gravitational-Wave Observatory (aLIGO) was shown to match the waveform predicted by general relativity for a pair of merging black holes (BBHs) \cite{TheLIGOScientific:2016src}. Such compact binary coalescence (CBC) events, also including merging black hole neutron star (NSBH) or neutron star pairs (BNS), are expected to be the most abundant sources, with detection rates between a few and tens per year \cite{0264-9381-27-17-173001, Abbott:2016nhf}. 

Detecting gravitational waves, and subsequently performing parameter estimation (PE) to infer the astrophysical parameters encoded in those waves is a key goal of gravitational-wave astronomy. Spin-induced precession of the binaries is a generic feature of gravitational waves emitted from CBC events,
and PE studies that neglect precession will ultimately suffer from (possibly large) systematic bias in the inferred parameter values \cite{PhysRevD.91.042003, PhysRevD.88.062001}.
However, including the effects of precession into template waveforms for PE carries a high computational cost associated with waveform generation and/or sufficiently sampling the astrophysical parameter space. The time it takes to complete an analysis scales (roughly) linearly with the waveform generation cost. There is therefore a
need to incorporate precession effects in PE studies in a computationally efficient way. Unless abated, computational costs are likely to increase
$(i)$ as more detailed physical effects are added to waveform models, e.g. higher-order modes, and $(ii)$ when in-band signals become longer as the detector's low-frequency sensitivity improves, making the detectors more sensitive to lower mass systems.

For parameter estimation studies, we are interested in computing the posterior probability density function (PDF)
\beq
p(\vec{\Lambda}| d)  = \frac{\mathcal{P}(\vec{\Lambda})\ \mathcal{L}(d | \vec{\Lambda})}{e(d)}\,, \label{e:posterior}
\eeq
on the set of model parameters $\vec{\Lambda}$, where $\mathcal{P}( \vec{\Lambda} )$ is the prior probability on the model parameters, $\mathcal{L}(d | \vec{\Lambda})$ is the likelihood of the data and $e(d)$ is known as the Bayesian ``evidence'' and describes the probability of the data given the model. The evidence is typically used for model selection and enters only as an overall scaling in parameter estimation. 

Assuming the detector data $d$ contains the GW signal $h(\vec{\Lambda}_{\tt true})$ and noise $n$, 
the log-likelihood function can be computed as
\begin{align}
\log\mathcal{L}(d|\vec{\Lambda}) &= - \frac{1}{2} (d - h(\vec{\Lambda}), d - h(\vec{\Lambda})) \, ,  \label{eq:likelihood} 
\end{align}
where $d = h(\vec{\Lambda}_{\tt true}) +  n$ and $(a, b)$ is an \textit{overlap} integral:
\beq
(d, h(\vec{\Lambda})) = 
4\Re\ \Delta f \sum_{k=1}^{L} \frac{\tilde d^*(f_k)\tilde h(f_k;\vec{\Lambda})}{S_{n}(f_k)}\, . \label{e:inner}
\eeq
Here $\tilde d(f_k)$ and $\tilde h(f_k;\vec{\Lambda})$ are the discrete Fourier transforms at frequencies $\{ f_k \}_{k=1}^L$ and $S_{n}(f_k)$ is the detector's noise power spectral density (PSD).
For a given observation time $T=1/\Delta f$ and detection frequency window $(f_{\tt high} - f_{\tt low})$ there are $L \sim 
{\tt int}\left( \left[ f_{\tt high} - f_{\tt low} \right] T \right)$ sampling points in (\ref{e:inner}).
When $L$ is large and $\vec{\Lambda}$ must be sampled extensively there are three bottlenecks: $(i)$ evaluation of the model at each $f_k$; $(ii)$ numerically computing the sum in the likelihood (\ref{eq:likelihood}); and $(iii)$ repeated evaluation of the likelihood.

These bottlenecks compound to escalate the cost of a typical parameter estimation analysis, even for otherwise fast-to-compute waveform models. 
Consider that a typical analysis can require computing several tens of millions of templates \cite{PhysRevD.91.042003}, and in principle these templates cannot be computed in parallel. Hence a single likelihood evaluation must be on the order of a millisecond for the 
PE analyses to be on the order of tens of hours. But this is often not the case.
Evaluating the closed-form frequency-domain waveform model known as IMRPhenomPv2 \cite{PhysRevLett.113.151101} - as implemented in the LIGO Algorithm Library \cite{LAL} - takes around half a second for a low mass systems starting from $20$Hz. These numbers imply PE run times on the order of six
months~\footnote{This time was the average of 100 waveform evaluations and overlap computations. For each evaluation we considered binary configurations with component masses of $1\,M_{\odot}$ and $4\,M_{\odot}$ and used random spin magnitudes and orientations on each iteration. The frequency resolution of the waveform was $\Delta f = 1/128$ Hz which assumes an in-band signal duration - rounded to the next-highest-power-of-two - of 128s from 20Hz, which is reasonable for such a binary configuration \cite{sathyaprakash2009physics}. All timing experiments, including this one, are performed using an Intel Xeon CPU with a 2.70GHz clock speed.}. Other commonly used waveform families incur similar or even higher computational costs. For example, the waveform family known as ``SEOBNRv2$\_$ROM'' \cite{0264-9381-31-19-195010} - a ``reduced order model'' of the aligned-spin waveform computed within the effective one body framework, and calibrated to numerical relativity - is only around a factor of four less expensive than IMRPhenomPv2. Conversely, the waveform family known as ``SEOBNRv3'' \cite{0264-9381-33-12-125025} - a precessing-spin waveform family computed within the effective one body framework, and calibrated to numerical relativity simulations - is around $170$ times \textit{more} expensive to compute than IMRPhenomPv2.

Reduced Order Modeling (ROM) is a promising technique for mitigating the computational cost of gravitational-wave parameter estimation. A ROM approach seeks to find a computationally efficient representation of the waveform model. If a set of $N < L$ basis elements can be found which accurately spans the continuum template space, it is possible to replace the overlap~(\ref{e:inner}) with a quadrature rule
containing only $N$ terms, reducing the overall cost by a factor of $L/N$.
This cost-reduction has been demonstrated in the context of gravitational waves from non-precessing CBCs \cite{PhysRevLett.114.071104}, but it was hitherto unclear that templates in the precessing case were also amenable to such \textit{linear} dimensional reduction. Here, linear refers to an approximation that is expressed as a linear superposition of basis elements. Non-linear dimensional reduction tools described in Refs.~\cite{Blackman2014,0264-9381-31-19-195010, Purrer:2015tud} are not directly applicable for \textit{compressed} overlap integrals.

A variety of ROM-type techniques have recently appeared in the GW literature \cite{PhysRevX.4.031006, PhysRevLett.114.071104, Caudill:2011kv, Field:2011mf, Blackman2014, 0264-9381-31-19-195010, Purrer:2015tud}.
We shall use a combination of the \textit{reduced basis method} and the \textit{empirical interpolation method}, whose favorable computational efficiency, ease-of-parallelization and numerical stability make them attractive candidates for tackling precessing waveform systems and other challenging models. The reduced basis method constructs a basis set of $N$ elements whose span reproduces the GW model within a specified accuracy. The empirical interpolation method then uses this model-specific basis to construct an $N$-point interpolant defined on the model space. Substituting the empirical interpolant representation into Eq.~\eqref{eq:likelihood} yields the \textit{reduced order quadrature} (ROQ)
rule~\cite{antil2012two,PhysRevLett.114.071104,canizares2013gravitational} which ultimately provides the performance gain of $L/N$. 

One of the caveats of the ROQ method is that, in order to realize the promised $L/N$ speedup, we must be able to directly evaluate the waveform model at special interpolation nodes in time or frequency. Typically, this means that the model is described by a closed-form expression.  Nevertheless, for other models, such as those described by differential equations, direct evaluation may be accomplished using \textit{surrogates}~\cite{0264-9381-31-19-195010, Purrer:2015tud, PhysRevX.4.031006,PhysRevLett.115.121102}. Although surrogate models have been constructed for
non-spinning~\cite{PhysRevLett.115.121102,PhysRevX.4.031006} and spin-aligned waveform models~\cite{0264-9381-31-19-195010, Purrer:2015tud}, it is not obvious that they can be (easily) constructed for precessing waveform models because surrogates rely on some form of high-dimensional fitting or interpolation. We return to this issue in the conclusion.

One of the main results of this paper is to apply the reduced basis and empirical interpolation methods 
to gravitational waveform models from CBCs with precessing spins. That this is possible should not be taken for granted. First, there are significant computational costs associated with long waveforms with multiple intrinsic parameters. To overcome this challenge we have developed and used a code called \texttt{greedycpp} that employs fast algorithms and possesses good scalability up to at least $32 \,, 000 $ cores~\cite{greedycpp,Antil2016}. Specially tailored parametric and frequency sampling strategies, discussed in Sec.~\ref{sec:TrainingSetQuadrature}, provide additional benefits. Second, although previous results 
show the existence of a compact basis for spin-aligned systems~\cite{PhysRevD.86.084046,Purrer:2015tud}, one may be worried that the complex waveform morphologies characteristic of precessing CBCs could result in a substantial increase in the basis size. This work demonstrates that there is no such increase.

Assuming that waveform generation and likelihood computation comprises the full cost of a PE study, 
we find theoretical speedup improvements between a factor of $4$ (for short BBH signals) and $300$ (for long BNS signals). The full range of speedup factors, which assumes that the signal is in-band starting at $20$Hz, is shown in Fig.~\ref{fig:speedup_runtimes}. Although we assume $f_{\tt low} = 20$Hz throughout this paper, we anticipate our speedup factors would increase (decrease) as $f_{\tt low}$ is lowered (raised) for a fixed value of the binary's masses (See Figure 1 of Ref.~\cite{PhysRevLett.114.071104}). If the entirety of the cost of parameter estimation is assumed to be the waveform and likelihood computations, we estimate a minimum run time of analyses from $6$ hours (for analyses on BBH signals up to $4$s in duration) to $12$ hours (for BNS/NSBH signals up to $128$s in duration). The speedup factors imply that run times without the ROQ could be on the order of 1 day to around 6 months using similar computer hardware and codes. We also show that modeling errors in the ROQ do not introduce additional systematic bias into PE, as shown in Sec.~\ref{sec:PE}.

The paper is outlined as follows. In Sec.~\ref{sec:prelim} we summarize the basics of ROQs for precessing gravitational waveform families. In Sec.~\ref{sec:strategy} we describe our strategies for working with a high-dimensional waveform model-space. In Sec.~\ref{sec:build_and_valid} we describe the results of running our basis-building pipeline and show the accuracy of the reduced basis and empirical interpolant. Using the \texttt{LALInference} library, in Sec.~\ref{sec:PE} we compare the accuracy of using the ROQ in a PE analysis to the Full likelihood function, for a simulated signal injected into recolored Gaussian noise designed to mimic early aLIGO data \cite{Berry:2014jja}. We also describe the speed up one could achieve by using the ROQ in PE analyses and we set a conservative performance benchmark for the run times of efficient PE codes. In Appendix~\ref{app:features} we describe a novel use of the reduced basis method as a diagnostic tool for waveform models and discuss its application to IMRPhenomPv2.

\section{Preliminaries}  
\label{sec:prelim}

\subsection{ROQ rules for precessing multi-modal gravitational wave models}
\label{sec:roq_rule}

A gravitational-wave strain signal $h(t)$ detected by a ground-based interferometer has the form 
\begin{align}
h(t;\vec{\Lambda}) = &F_{+}\left(\text{ra},\text{dec},\psi,r\right) h_{+}(t;\phi_c,t_c,\vec{\lambda}) + \nonumber \\
                     &F_{\times}\left(\text{ra},\text{dec},\psi,r\right) h_{\times}(t;\phi_c,t_c,\vec{\lambda}) \, ,
\end{align}
where the antenna patterns $F_{(+,\times)}$ project the gravitational wave's 
$+$- and $\times$-polarization states, $h_{(+,\times)}$, into the detector's frame. 
The antenna patterns are functions of variables which 
specify the orientation of the detector with respect to the binary:
the distance to the source ($r$) as well as the right ascension ($\text{ra}$), 
declination ($\text{dec}$) and polarization ($\psi$) angles. These four variables,
along with the coalescence time ($t_c$) and its orbital phase at coalescence ($\phi_c$), describe the signal's
dependence on parameters that have a trivial effect on the waveform's amplitude and phase.
We shall use $\vec{\lambda}$ to 
denote the signal's dependence on parameters that have a non-trivial effect on the waveform's amplitude and phase, such as its masses, spin magnitude and spin orientation \footnote{Note that we refrain from discussing ``intrinsic'' and ``extrinsic'' parameters because for precessing systems, extrinsic parameters like the binary's orbital inclination can produce non-trivial effects in the waveform's amplitude and phase and so they do not simply enter as scaling factors as in non-precessing systems. Our ROQ rule is trained over the subset of parameters $\vec{\lambda}$
but applies to the full set $\vec{\Lambda}$ (cf. Sec.~\ref{sec:PeskyParams}).}.
The strain, and consequently the likelihood \eqref{eq:likelihood}, 
depends on the full set of parameters
$\vec{\Lambda} = \{\text{ra},\text{dec},\psi,r,t_c,\phi_c, \vec{\lambda} \}$.

When discussing waveform models, it is common practice to first introduce a
complex gravitational wave strain
\begin{align}
h_{+}(t;\phi_c,t_c,\vec{\lambda}) &- {\mathrm i} h_{\times}(t;\phi_c,t_c,\vec{\lambda}) \nonumber \\
& =  \sum_{\ell=2}^{\infty} \sum_{m=-\ell}^{\ell} h^{\ell m}(t;\phi_c,t_c,\vec{\lambda}) {}_{-2}Y_{\ell m} \, ,
\end{align}
which is subsequently decomposed into a basis of spin-weighted spherical harmonics. Most gravitational waveform 
models make predictions for the modes $h^{\ell m}(t;\vec{\lambda})$, from which a model 
of what a noise-free detector 
records, $h(t;\vec{\Lambda})$, is readily recovered.

The remainder of this subsection sketches the steps leading to the reduced order quadrature rule.
To build computationally efficient approximations 
to \eqref{eq:likelihood}, we work directly
with the Fourier transform of the strain 
\begin{align}
\tilde{h}(f;\vec{\Lambda}) & = \int_{-\infty}^{\infty} h(t;\vec{\Lambda}) 
                               \mathrm{e}^{2 \pi \mathrm{i} f t} dt \nonumber \\
& = F_{+} \tilde{h}_{+}(f;\phi_c,t_c,\vec{\lambda})  
  + F_{\times}\tilde{h}_{\times}(f;\phi_c,t_c,\vec{\lambda}) \nonumber \\
& = \mathrm{e}^{-2 \pi \mathrm{i} f t_c} \left[ F_{+} \tilde{h}_{+}(f;\phi_c,0,\vec{\lambda})  
  + F_{\times}\tilde{h}_{\times}(f;\phi_c,0,\vec{\lambda}) \right] \nonumber \\
\label{eq:hoff}
\end{align}
where the antenna pattern's arguments are omitted for brevity. 
The last equality follows from $h(t;t_c) = h(t-t_c;0)$, 
as a non-zero coalescence time $t_c$ simply offsets the 
signal's time-of-arrival. Because $\tilde{h}_{(+,\times)}$ enters linearly into 
$(d, h)$ and quadratically into $(h, h)$, 
one of the goals of this paper is to build (temporarily 
focusing on the model's internal parameterization $\vec{\lambda}$)
an approximation 
\newpage 
\begin{widetext}
\begin{subequations}\label{eq:EIM}
\begin{gather}
\tilde{h}_{\text{A}}(f_i;\vec{\lambda}) \approx  \sum_{j=1}^{N_{\tt L}} B_j (f_i) \tilde{h}_{\text{A}}(F_j;\vec{\lambda})\,, 
\quad \text{with}\, \text{A} \in \{ +,\times \}  \label{eq:EIM_lin}\,,\\
 \Re  \left[ \tilde{h}_{\text{A}}(f_i;\vec{\lambda})\tilde{h}_{\text{B}}^*(f_i;\vec{\lambda}) \right]
\approx \sum_{k=1}^{N_{\tt Q}} C_k (f_i) 
     \Re \left[ \tilde{h}_{\text{A}}(\mathcal{F}_k;\vec{\lambda})\tilde{h}_{\text{B}}^*(\mathcal{F}_k;\vec{\lambda}) \right] \,,
\quad \text{with}\, \text{A},\text{B} \in \{+,\times\} \label{eq:EIM_quad} \,,\
\end{gather}
\end{subequations}
\end{widetext}
that accurately approximates both the polarization states and their products. 
Here the labels $\text{A}$ and $\text{B}$ take the values $(+,\times)$, 
$\{B_j\}_{j=1}^{N_{\tt L}}$ is the reduced basis (RB) for the polarizations and $\{C_k\}_{k=1}^{N_{\tt Q}}$ is the RB for 
the real part of all possible products of the polarizations. Notice that in Eq.~(\ref{eq:EIM_lin}) $\tilde{h}_{\text{+}}$ and $\tilde{h}_{\times}$ share the \textit{same} basis $\{B_j\}_{j=1}^{N_{\tt L}}$. Similarly the approximation to the products in Eq.~(\ref{eq:EIM_quad}) $\tilde{h}_{\text{+}}\tilde{h}_{\text{+}}^{*}$, $\tilde{h}_{\times}\tilde{h}_{\times}^{*}$ and $\Re \tilde{h}_{\text{+}}\tilde{h}_{\times}^{*}$ also share a basis $\{C_k\}_{k=1}^{N_{\tt Q}}$.
The values $\tilde{h}_{\text{A}}(\vec{\lambda}; F_j)$ are 
evaluations of the $\text{A}$-polarization states at the \textit{empirical interpolation nodes} $\{F_j\}_{j=1}^{N_{\tt L}}$. The location of these nodes are uniquely selected
to yield accurate interpolation with the set of basis vectors $\{B_j\}_{j=1}^{N}$. 
Similarly, polarization products $\tilde{h}_{\text{A}}(\mathcal{F}_k;\vec{\lambda})\tilde{h}_{\text{B}}^*(\mathcal{F}_k;\vec{\lambda})$ are evaluated at a set of empirical interpolation nodes
$\{\mathcal{F}_k\}_{k=1}^{N_{\tt Q}}$, which are distinct from $\{F_j\}_{j=1}^{N_{\tt L}}$. 
The approximation (\ref{eq:EIM}) is known as an \textit{empirical interpolant}, and its substitution into
into~(\ref{e:inner}) yields a \textit{reduced order quadrature} (ROQ) rule. The empirical interpolant constitutes a ROM of the waveform family.
Sec.~\ref{sec:algorithms} describes the algorithms we use to build~\eqref{eq:EIM}. As described in Sec.~\ref{sec:PeskyParams}, 
with the exception of $t_c$
the approximation~\eqref{eq:EIM} automatically applies to the model's full parameterization $\vec{\Lambda}$
despite being built for the subset of internal model parameters $\vec{\lambda}$. 
In many of the expressions which follow we shall use $\vec{\Lambda}$ to denote 
the full parameter vector but with $t_c$ explicitly separated off.

We break the likelihood into those pieces which 
we can approximate using~\eqref{eq:EIM}
\begin{widetext}
\begin{align} \label{eq:ROQ_likelihood1}
2 \log\mathcal{L}  & =
2 (d,h) - (h,h)  - (d,d)  \nonumber \\
& 
= 2 F_+ (d,h_+) + 2 F_\times (d,h_\times) 
- \left| F_+ \right|^2 (h_+,h_+)
- \left| F_\times \right|^2 (h_\times,h_\times)
- 2 F_+ F_\times (h_+,h_\times) - (d,d) \nonumber \\
& 
\approx 2 F_+ (d,h_+)_{\text{ROQ}} + 2 F_\times (d,h_\times)_{\text{ROQ}}
- \left| F_+ \right|^2 (h_+,h_+)_{\text{ROQ}}
- \left| F_\times \right|^2 (h_\times,h_\times)_{\text{ROQ}}
- 2 F_+ F_\times (h_+,h_\times)_{\text{ROQ}}  - (d,d) \nonumber \\
& 
= 2 \log\mathcal{L}_{\text{ROQ}} \, .
\end{align}
\end{widetext}
The linear
\begin{subequations}\label{eq:ROQ_part1}
\begin{eqnarray}
(d,h_{\text{A}}(\vec{\lambda}) )_{\text{ROQ}}
&\approx& \sum_{j=1}^{N_{\tt L}}\omega_j(t_c) \tilde{h}_{\text{A}}(F_j;\vec{\lambda}) \,,\\
\omega_j(t_{c}) &=& 4\Re\ \Delta f \sum_{i=1}^L \frac{ \tilde{d}^*(f_i) B_j (f_i)}{ S_n(f_i)} \mathrm{e}^{-2 \pi \mathrm{i} t_{c} f_i } \,
\end{eqnarray}
\end{subequations}
and quadratic 
\begin{subequations}\label{eq:ROQ_part2}
\begin{eqnarray}
(h_{\text{A}}(\vec{\lambda}),h_{\text{B}}(\vec{\lambda}) )_{\text{ROQ}}
&\approx& \sum_{k=1}^{N_{\tt Q}}\psi_k \tilde{h}_{\text{A}}(\mathcal{F}_k; \vec{\lambda})\tilde{h}^{*}_{\text{B}}(\mathcal{F}_k;\vec{\lambda}) \,,\\
\psi_k &=& 4\Re\ \Delta f \sum_{i=1}^L \frac{C_k (f_i)}{ S_n(f_i)} \,,
\end{eqnarray}
\end{subequations}
ROQ rules are straightforward to derive: simply substitute the relevant approximations~\eqref{eq:EIM}
into each of the five overlaps~\eqref{e:inner} appearing after the second equality in~\eqref{eq:ROQ_likelihood1}. 
Notice that the \textit{data-dependent} weights
$\omega_j$ are composed of full overlaps~\eqref{e:inner} 
between all the basis elements and the whitened data.
While the weights $\psi_k$ in the quadratic ROQ rule do not depend on the data stream $\tilde{d}(f)$, 
they do depend on the power spectral density
$S_n(f)$ which, for the most realistic scenarios, is experimentally estimated.
The next section describes 
our approach for the dependence of \eqref{eq:ROQ_part1} on $t_c$.
Generation of both flavors of weights comprise the ROQ \textit{startup cost}.
Once the weights are known, computing the ROQ likelihood 
only requires $N_{\tt L}+N_{\tt Q}$ terms (hence only $N_{\tt L}+N_{\tt Q}$ waveform model evaluations), 
thereby reducing the cost of (\ref{e:inner}) by a factor of $L/(N_{\tt L}+N_{\tt Q})$.

Using the definition of the weights (\ref{eq:ROQ_part1}b,\ref{eq:ROQ_part2}b) and the 
reality of the basis set $\{C_k\}_{k=1}^{N_{\tt Q}}$, expression \eqref{eq:ROQ_likelihood1}
can be written in a convenient form for numerical implementation as
\begin{align} \label{eq:roq_simple}
2 \log\mathcal{L}(d|\vec{\Lambda})_{\text{ROQ}} + (d,d) & =  \nonumber \\
& \hspace{-55 pt} 2\Re \sum_{j=1}^{N_{\tt L}} \omega_j (t_{c}) \tilde{h}(F_j;\vec{\Lambda}) 
- \sum_{k=1}^{N_{\tt Q}} \psi_j \tilde{h}(\mathcal{F}_k;\vec{\Lambda}) \tilde{h}^*(\mathcal{F}_k;\vec{\Lambda}) \, .
\end{align}
Compared to the usual likelihood expression~\eqref{eq:likelihood} 
using the typical overlap~\eqref{e:inner},
\begin{align}
2 \log\mathcal{L}(d|\vec{\Lambda}) + (d,d) & =  \nonumber \\
& \hspace{-75 pt} 2 \Re\ \sum_{l=1}^{L} \frac{4  \Delta f \tilde d^*(f_l)}{S_{n}(f_l)} \tilde{h}(f_l;\vec{\Lambda})
-  \sum_{l=1}^{L} \frac{4  \Delta f  }{S_{n}(f_l)} \tilde{h}(f_l;\vec{\Lambda}) \tilde{h}^*(f_l;\vec{\Lambda}) \, ,
\end{align}
shows the ROQ rule to be similar to the standard evaluation pattern, 
thereby allowing existing codes to easily implement these tools. The simplified expression~\eqref{eq:roq_simple}
necessarily requires our basis to permit approximations of the form~\eqref{eq:EIM}. In
particular, had we instead built a separate basis for each polarization and product piece, we would
have been forced to retain all five terms originally present in Eq.~\eqref{eq:ROQ_likelihood1}.

\subsection{Trivial and nontrivial parameters}
\label{sec:PeskyParams}
 
Certain parameters need not be included in
the training of the ROM representation~\eqref{eq:EIM}.
In practice, this means 
we can explicitly set these ``neglected" parameters to
a fixed constant.
In most cases this is the correct thing to do.
The distance to the source, for example,
affects the strain as multiplication by an overall
constant. Consequently,
if $\tilde{h}(f;r = 1, \ldots)$ can be 
accurately integrated with an ROQ rule
then so can $\tilde{h}(f;r \neq 1, \ldots)$.  
Simply evaluate Eq.~\eqref{eq:roq_simple} at 
the desired value of $r$.
Sky position, orientation and
orbital phase at coalescence affect
the strain in a similar, frequency-independent manner\footnote{Conversely, 
the inclination angle which is normally considered 
``extrinsic'' in non-precessing models is ``promoted'' to an
intrinsic parameter in precessing models because it is frequency 
dependent. As such, the extension of inclination to precessing 
systems is included in the parameter vector $\vec{\lambda}$ in 
Eq.~\ref{eq:EIM}}.

A notable exception is the signal's arrival time.
Our approach for the dependence of \eqref{eq:ROQ_part1}
on $t_c$ follows Ref.~\cite{canizares2013gravitational}: 
a unique set of ROQ weights is constructed for
$n_c$ equally spaced values of $t_c$
sampling the interval $[ t_{\tt trigger} - W, t_{\tt trigger} + W]$,
where an estimate for the time window $W$ centered around the coalescence 
time $t_{\tt trigger}$ is given by the GW search pipeline.
Instead of using nearest neighbor interpolation,
as was done in Ref.~\cite{canizares2013gravitational},
we use spline interpolation to evaluate the weights 
at arbitrary values of $t_c$. 
Since the weights $\omega_j(t_{c})$ are smooth functions of $t_c$
they are well suited for higher-order interpolation. 
This means, as compared to nearest neighbor interpolation, 
significantly higher accuracies and/or use significantly 
smaller values of $n_c$ are achieved with a spline.

When data is recorded at multiple detectors, inference is carried out using a model
whose parameterization is again given by Eq.~\eqref{eq:hoff}. 
The ROQ works the same as before, so long as 
one takes into account the possible time-of-arrival offsets 
when computing $\omega_j(t_{c})$. To handle this, we pad the time window
estimates $W$ by $\pm 26$ms, which is the duration required for a classical 
gravitational wave to travel from the Earth's geocenter to any conceivable 
earth-based GW detector. This allows the $t_c$-dependent ROQ weights 
to be applicable for all network detectors. 

\subsection{Numerical algorithms}
\label{sec:algorithms}

The reduced order quadrature rule is trained on a dense training set of waveforms using the
algorithms of Refs.~\cite{antil2012two,canizares2013gravitational,PhysRevLett.114.071104} which have been 
implemented in C++ and parallelized with message passing interface~\cite{greedycpp,Antil2016}. First, on this training set we apply a greedy algorithm (see algorithm $1$ of Ref.~\cite{canizares2013gravitational}) to construct a nearly optimal reduced basis for the waveform family~\cite{Binev11,Field:2011mf}. The algorithm proceeds from a linear basis constructed from $i$ waveforms already chosen. For each training set waveform, we compute the best possible approximation given as a linear combination of the basis elements. The approximate waveform with the largest error is added to the basis as its $i + 1$ element. Next, given $N$ basis elements we find the $N$ uniquely determined empirical interpolation nodes with another, different greedy strategy~\cite{Maday_2009,chaturantabut2010nonlinear}. Our implementation of the empirical interpolation method uses the modification suggested by Ref.~\cite{antil2012two} which reduces the overall cost from $\mathcal{O}\left(N^4\right)$ to $\mathcal{O}\left(N^3\right)$ (see algorithm $2$ of Ref.~\cite{canizares2013gravitational}).

Out-of-training-set validation is carried out by computing the approximation error of randomly sampled waveforms. Typically we use $\approx 10^7$ random samples, which trivially parallelizes with OpenMP within each compute node. We record errors larger than $10^{-6}$, adding these waveforms back into the original training set. On this enriched set we reapply the greedy basis building algorithm, thereby producing a more accurate basis. The ROQ building procedure, as just described, is largely automated~\cite{greedycpp}. 

\subsection{Phenomenological model for precessing inspiral, merger and ringdown waveforms}
\label{sec:phenomp}

Waveform models are available for a variety of binary configurations. The most general models include configurations in which the individual spin angular momenta $S_i$ of the compact objects are allowed to be misaligned with the orbital angular momentum $\hat{L}$ of the binary. This spin misalignment is the source of more complex binary dynamics which causes the orbital plane, as well as the individual spins, to precess~\cite{PhysRevD.49.6274, Kidder:1995zr,ossokine2015comparing}. 
Depending on the relative orientation between the source and the observer, mild to strong amplitude and phase
modulations are observed in the GW signal (see, e.g., Ref.~\cite{Schmidt:2012rh} for an illustration). Only recently have precessing waveform models describing an approximate inspiral-merger-ringdown (IMR) signal become available~\cite{PhysRevLett.113.151101, Pan:2013rra}.
 
The waveform model used in this paper is a phenomenological waveform model known as IMRPhenomPv2 as implemented in the LIGO Algorithm Library (LAL) \cite{LAL}.
This model describes an approximate IMR signal of precessing binary black holes by appropriately rotating the waveforms of an aligned-spin system by means of Euler rotations into the modes exhibited by a precessing system~\cite{PhysRevLett.113.151101}. 
Schematically, this ``twisting up'' procedure may be expressed as~\cite{Schmidt:2012rh}:
\begin{equation}
\label{eq:twist}
h^\mathrm{prec}_{\ell m} = \sum_m \mathbf{ R}_{\ell m} h^\mathrm{aligned}_{\ell m}, 
\end{equation}
where $\mathbf{R}_{\ell m}$ denotes the operator which encoded the relevant Euler rotations.
This requires three main ingredients: an accurate aligned-spin model, a description of the orbital precession dynamics and a prediction for the spin and mass of the resulting black hole remnant. 

The underlying aligned-spin IMR waveform model is IMRPhenomD~\cite{Husa:2015iqa, Khan:2015jqa}, an aligned-spin waveform model which provides only the $(2, |2|)$-modes of the GW signal. Its inspiral portion has been extensively calibrated to effective-one-body waveforms~\cite{Taracchini:2012ig}, and the merger part to Numerical Relativity (NR) waveforms for binary configurations with dimensionless spin magnitudes between -0.95 and 0.98 and mass ratios between 1 and 18. 

To model the precession of the orbital plane, analytic post-Newtonian (PN) expressions through second post-Newtonian (PN) order in spin-orbit terms~\footnote{The orbital angular momentum, however, uses a 2PN expression without any contribution from the spin terms.} are used~\cite{PhenomPv2}. The ``twisting-up'' procedure Eq.~\ref{eq:twist} results in a precessing waveform model which contains all $\ell=2$ waveform modes. However, the absence of the $m=0$ and $m = \pm 1$ modes in IMRPhenomD leads to approximate precessing modes. The spin and mass of the final black hole are obtained from fits to NR data~\cite{PhenomPv2}. We note that IMRPhenomPv2 has not been directly calibrated against precessing NR waveformsand does not include any tidal effects.

\begin{figure}
    \centering
    \def\svgwidth{1.2 \columnwidth}
    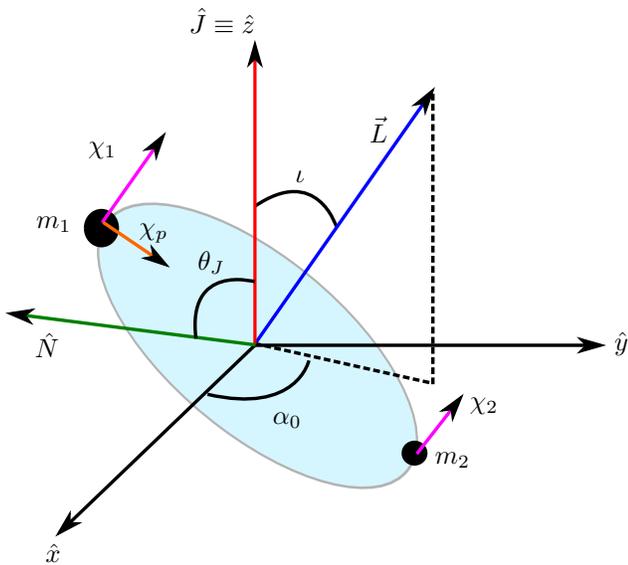
    \caption{The $J$-aligned source frame of a precessing binary. IMRPhenomPv2 uses a single precessing spin approximation to describe the inspiral, merger and ringdown and is described by the parameter vector $\vec{\lambda} = (m_1, m_2, \chi_{1}, \chi_{2}, \chi_p, \theta_J, \alpha_0)$, with $\hat{N}$ in the xz-plane. Here, $\chi_{1}$ and $\chi_{2}$ are the spin components that lie parallel to $\vec{L}$ on the heavier ($\chi_{1}$) and lighter ($\chi_{2}$) compact object; and the perpendicular spin parameter $\chi_p$ is the spin component that lies in the orbital plane and is associated with the heavier body $m_1$.
\label{fig:frame}
}
\end{figure}

To compute the gravitational-wave polarizations $h_+$ and $h_{\times}$, it is convenient to adopt a time-independent Cartesian source frame attached to the binary.  
For aligned-spin binaries, a common choice is a coordinate frame such that $\hat{L} \equiv \hat{z}$. In the case of precession, however, $\hat{L}$ evolves with time, but the direction of the total angular momentum, $\vec{J} = \vec{L} + \vec{S}_1 + \vec{S}_2$, stays approximately fixed during the binary's orbital evolution. A natural choice for the binary source frame therefore is a Cartesian coordinate system, where $\hat{J}$ at some reference gravitational-wave frequency $f_{\tt ref}$ defines the z-axis. This source frame is depicted in Fig.~\ref{fig:frame}.

In the following, we denote the angle between the line-of-sight $\hat{N}$ and $\hat{J}$ by $\theta_J$. The relative orientation between $J$ and the GW detector significantly affects the morphology of a precessing signal. The parameter $\theta_J$ represents the natural generalization of the inclination of the orbital plane and is therefore an important parameter to be taken into account when building the ROM/ROQ~\footnote{Alternatively, one could build an ROQ for the individual modes.}. Another relevant parameter is the azimuthal orientation of the orbital angular momentum $L$ at the reference gravitational-wave frequency $f_{\tt ref}$ denoted by $\alpha_0$. The evolution of $\alpha(t)$ encodes the precession of $L$ around $J$ and is thus often referred to as the ``precession angle''~\cite{PhysRevD.49.6274}. Together the Euler angles $\alpha(t)$ and $\iota(t)$ (defined in Fig.~\ref{fig:frame}) ``twist-up'' the non-precessing carrier model IMRPhenomD.
The other model parameters are the component masses, $m_1$ and $m_2$ with $m_1 \geq m_2$, the dimensionless spin magnitudes projected onto the orbital angular momentum $\hat{L}$, $\chi_{1}$ and $\chi_{2}$,
and one ``effective'' precessing spin parameter $\chi_p$~\cite{PhysRevD.91.024043} defined as
\begin{equation}
\label{eq:chip}
\chi_p = \frac{\max{(A_1 m_1^2 \chi_{1\perp}, A_2 m_2^2 \chi_{2\perp}})}{A_1 m_1^2},
\end{equation}
where $A_1 = 2 + 3m_2/2m_1$, $A_2 = 2 + 3m_1/2m_2$ and $\chi_{i\perp}$ are the magnitudes of the spin vectors perpendicular to $\hat{L}$, i.e., the spin projections into the orbital plane. 
The motivation for this choice of effective parameterization is the following: In general, a precessing binary can have up to four spin components orthogonal to L, which are \textit{all} the source of precession. However, these can be combined efficiently into a single precessing spin parameter, $\chi_p$, which when applied to the heavier body ($m_1$), captures the average precession exhibited by the system with all four in-plane spin components~\cite{PhysRevD.91.024043}. 

The relevant IMRPhenomPv2 parameters are given by $\vec{\lambda} = (m_1, m_2, \chi_{1}, \chi_{2}, \chi_p, \theta_J, \alpha_0)$. Other parameters that enter as an overall scaling, such as distance to the source or its position in the sky, are omitted in the waveform model itself as these can be included trivially. The model parameters in the $\hat{J}$-aligned source binary frame are shown in Fig.~(\ref{fig:frame}) which is adapted with permission from Ref.~\cite{PhysRevD.91.024043}.

Various simplifying assumptions have been made in the current implementation\footnote{The results of this paper use the IMRPhenomPv2 model as implemented in the LAL with a git hash of a50aca13b97412999fad03a073a6a5b319fd5bc4.} of the IMRPhenomPv2 waveform model. One is that $\hat{J}$ is kept constant, and that the angle between $\hat{L}$ and $\hat{J}$ is small. We therefore do not expect that IMRPhenomPv2 accurately models precessing cases where $J \sim 0$. Such cases, observed for highly anti-aligned spins with moderate mass ratios and only a small value of $\chi_{p}$, are known as transitional precession as $\hat{J}$ undergoes a ``flip'' and completely changes its orientation~\cite{PhysRevD.49.6274}. We also do not expect waveforms of systems with higher mass ratios and large values of $\chi_{p}$ to be modeled accurately by IMRPhenomPv2 as the angle between $L$ and $J$ can be large for such cases. However, for some of these cases the model may still produce acceptable results, but detailed checks across the parameter space have not yet been performed.

\begin{table*}
\centering
\begin{tabular}{c | c | c c | c | c | c c | c c |c }

\multirow{2}{*}{Case} 
& \multirow{2}{*}{Build strategy} 
& \multicolumn{2}{|c|}{$f$ (Hz)}
& \multirow{2}{*}{Waveform duration T} 
& \multirow{2}{*}{$\Delta f$ (Hz)} 
& \multicolumn{2}{|c}{$\mathcal{M}$ ($M_{\odot}$)}
& \multicolumn{2}{|c|}{Basis size}
& \multirow{2}{*}{Speedup} \\

        &  
	& Min \!& \!\!Max\! \!
        &
        &
        & Min \!& \!\!Max\! \!
        & Linear \!& \!\!Quadratic\\
\hline
A     & Enriched greedy
        & $20$ & $1024$
        & $1.5\text{s} \leq T \leq 4\text{s}$
        & 1/4
        & 12.3 & 23
        & 300 & 197
        & 8 \\
B     & Enriched greedy
        & $20$ & $1024$
        & $3\text{s} \leq T \leq 8\text{s}$
        & 1/8
        & 7.9 & 14.8
        & 388 & 278
        & 12\\
C     & Enriched greedy
        & $20$ & $2048$
        & $6\text{s} \leq T \leq 16\text{s}$
        & 1/16
        & 5.2 & 9.5
        & 360 & 233
        & 54\\     
D     & Enriched greedy
        & $20$ & $2048$
        & $12\text{s} \leq T \leq 32\text{s}$
        & 1/32
        & 3.4 & 6.2
        & 524 & 254
        & 83\\    
E     & Enriched greedy
        & $20$ & $2048$
        & $23.8\text{s} \leq T \leq 64\text{s}$
        & 1/64
        & 2.2 & 4.2
        & 749 & 270
        & 127\\
F     & Enriched greedy
        & $20$ & $4096$
        & $47.5\text{s} \leq T \leq 128\text{s}$
        & 1/128
        & 1.4 & 2.6
        & 1253 & 487
        & 300\\\hline
A$^{\prime}$     & A scaled
        & $10$ & $512$
        & $3\text{s} \leq T \leq 8\text{s}$
        & 1/8
        & 24.6 & 46
        & 300 & 197
        &8\\
B$^{\prime}$     & B scaled
        & $10$ & $512$
        & $6\text{s} \leq T \leq 16\text{s}$
        & 1/16
        & 15.8 & 29.6
        & 288 & 278
        &12\\
C$^{\prime}$     & C scaled
        & $10$ & $1024$
        & $12\text{s} \leq T \leq 32\text{s}$
        & 1/32
        & 10.4 & 19
        & 360 & 233
        &54\\     
D$^{\prime}$     & D scaled
        & $10$ & $1024$
        & $23.8\text{s} \leq T \leq 64\text{s}$
        & 1/64
        & 6.8 & 12.4
        & 524 & 254
        &83\\    
E$^{\prime}$     & E scaled
        & $10$ & $1024$
        & $47.5\text{s} \leq T \leq 128\text{s}$
        & 1/128
        & 4.4 & 8.4
        & 749 & 270
        &127\\
F$^{\prime}$     & F scaled
        & $10$ & $2048$
        & $95\text{s} \leq T \leq 256\text{s}$
        & 1/256
        & 2.8 & 5.2
        & 1253 & 487
        &300
\end{tabular}
\caption{Regions in parameter- and mass-frequency- space in which we build a distinct ROM. Each case corresponds to an overlapping region in chirp-mass ($\mathcal{M}$) space. In all cases the bases/interpolants are valid in the mass-ratio interval $1 \leq q \leq 9$ which is within IMRPhenomPv2's calibration-range~\cite{Khan:2015jqa}. This range in mass ratio allows us to describe BNS systems, NSBH systems and BBH systems. Additionally, we impose the constraint on the component masses $m_1 \geq m_2 \geq 1\,M_{\odot}$.
For each case, we limit the magnitudes of the spin-related parameters ($\chi_{\text{1}}, \chi_{\text{2}}, \chi_{p}$) to lie within the range $(-0.9,-0.9, 0) \leq (\chi_{\text{1}}, \chi_{\text{2}}, \chi_{p}) \leq (0.9, 0.9, 0.9)$ and we use the full range for the spin angles: $(0,0)\leq (\theta_J, \alpha_0) \leq (\pi, 2\,\pi)$.
Cases A$^{\prime}$- F$^{\prime}$ show how the mass and frequency ranges of 
cases A-F can be scaled (See Sec.~\ref{sec:translate}) to $10$Hz without any additional computational effort.
\label{tab:basis_ranges}
}
\end{table*}

\section{Strategies for building high dimensional GW ROMs}
\label{sec:strategy}

Previous work \cite{PhysRevD.86.084046, Caudill:2011kv, PhysRevX.4.031006, Blackman2014, 0264-9381-31-19-195010, Purrer:2015tud, PhysRevLett.114.071104} on constructing reduced bases of waveform models have considered waveforms described by only a few \textit{intrinsic} parameters or short signals.
The IMRPhenomPv2 waveform family is described by seven parameters
and the waveform morphologies are inherently  more complex than in the non-precessing case. The increase in the size of the parameter space, together with the greater variety of waveform morphologies, means that constructing a faithful training space is more difficult than in previous work. 

Another concern has to do with the fact that we would like the ROQ to be useful for a very large range of astrophysically relevant parameters; from binary neutron stars with a total mass of around $2\,M_{\odot}$ to binary black holes with total masses of several tens of solar masses. The signals associated with these different ends of the mass spectrum have very different in-band durations.

In this section we describe our strategy for dealing with these issues as they relate to populating a faithful training set. We also provide a short review of approaches used in previous work. 

\subsection{Mass and frequency partitions}

We would like our ROQ to be valid for BNS, NSBH and BBH systems with as few basis elements as possible. In addition, we want to be able to exploit the lowest sensitive frequency of the detectors. To ensure these conditions are met, we find it useful to partition the mass-space into (overlapping) regions in chirp mass. These overlapping regions are defined by
\beq
\mathcal{M}(T=2^{n+1}s) \leq \mathcal{M} \leq 1.2\,\mathcal{M}(T=2^{n}s) \,,
\eeq
where $T$ is the waveform duration \cite{LAL} from $20$Hz, $\mathcal{M}=(m_1 m_2)^{3/5}/(m_1+m_2)^{1/5}$ is the chirp mass, which specifies the waveform duration to leading order, and $q = m_1/m_2 \geq 1$ is the mass ratio chosen between 1 and 9. To interpret $\mathcal{M}$ as a function of time (and vice versa) we build an interpolant of $\mathcal{M}(T)$ 
using the LAL function \texttt{SimIMRSEOBNRv2ChirpTimeSingleSpin}.
We compute the signal duration for a given chirp mass, fixing the spins to be maximally prograde and mass ratio to be $9$, which produces the longest inspirals \cite{misner1973gravitation}. We consider the following powers of $2$; $n=2,\, 3,\, \ldots,\, 6$, corresponding to regions in $\mathcal{M}$-space describing signals with durations; $1.5\text{s} \leq T \leq 4\text{s}$; $3\text{s} \leq T \leq 8\text{s}$; $6\text{s} \leq T \leq 16\text{s}$; $12\text{s} \leq T \leq 32\text{s}$; $23.8\text{s} \leq T \leq 64\text{s}$; $47.5\text{s} \leq T \leq 128\text{s}$. The union of the overlapping regions in chirp mass capture binary systems with signal-durations between slightly less than $2$s up to $128$s starting from $20$Hz.   The upper frequencies for the cases in Table \ref{tab:basis_ranges} correspond to the maximum-over-configuration ringdown frequency, rounded the next-highest-power-of-two.

Our particular choices have been guided by the expectation that, typically, a stochastic sampler will stay confined to a given partition or two. Future improvements to the ROQ method presented here, and more generally ROM building, may find different partition strategies to work better. Notice that the finer we make our mass partition the fewer basis will be needed in each partition and hence yields greater ROQ compression. Finer mass partitions also reduce the \textit{offline} cost associated with building the basis in a given partition. On the other hand, if we add up all the basis from all the partitions we should expect to find this total to be larger than a corresponding basis resulting from one large partition of equivalent extent. Both small~\cite{cannon2010singular} and large~\cite{Field:2011mf} partitions have been considered in other contexts.

\subsection{Ranges in mass ratio and spin}

Unlike our treatment of the chirp mass, we do not use any special partitioning strategy for the $6$ remaining parameters. Table~\ref{tab:basis_ranges}
and its caption summarizes the default parameter intervals used for the mass ratio ($q$) and the spin-related parameters ($\chi_{\text{1}}, \chi_{\text{2}}, \chi_{p}, \theta_J, \alpha_0$).

Since we are working with internal IMRPhenomPv2 parameters we have to impose constraints on some of the model's spin-related parameters~\cite{PhenomPv2}. These constraints eliminate unphysical systems with spins above the Kerr limit. The original physical BH binary can have in-plane spin components $\chi_{i,p}$ on either BH $i=1,2$. The spins must satisfy the Kerr limit on each BH: $\chi_{i,p}^2 + \chi_{i}^2 \leq 1$. Since
the model's effective precessing spin satisfies $\chi_{p} \leq \max [\chi_{1,p}, W(q) \chi_{2,p}]$ with $W(q) = \frac{3q+4}{4q^2 + 3q}$, in practice simply excluding $\chi_{p}^2 + \chi_{1}^2 \geq 1$ is good enough. 

We have had to place one additional restriction on the spins. Specifically, we exclude the region where and $\chi_{1} \leq 0.4 - 7\eta$. This constraint arises because the model exhibits non-smooth, rapidly changing behavior with parametric variation thereby precluding the existence of an accurate, sparse basis. We describe this problematic region in Appendix~\ref{app:features}.

\subsection{Deterministic and random sampling of the parameter space}
\label{section:TS}

Previous work~\cite{PhysRevD.86.084046, Caudill:2011kv, PhysRevX.4.031006, 0264-9381-31-19-195010, Purrer:2015tud, Blackman2014} has shown that a good strategy for sampling in the mass space is to sample uniformly in $\mathcal{M}^{3/5}$ as this is the leading order mass term that enters into the waveform phasing. It has also been observed that the basis elements are preferentially selected from the boundary of the parameter space, suggesting an efficient training set would overpopulate these regions. Additionally, the authors of Ref.~\cite{Blackman2014} considered a random greedy sampling strategy for precessing waveforms, parametrized by phase in a co-precessing frame. In this framework, a new training set is randomly generated at each iteration of the greedy algorithm thereby allowing for an effectively greater number of training waveforms. This strategy was motivated by the cost of storing the training set in memory which we overcome by using a parallelized code. 

For cases A-C in Table~\ref{tab:basis_ranges}, we find that using just $8$ sample points on a uniform grid in $\mathcal{M}^{3/5}$ and $\eta$ and a uniform grid of $8$ points in each of the remaining five spin-parameters yields a reasonably accurate basis. For the more challenging cases D-F in Table~\ref{tab:basis_ranges}, we increase our set to $64$ sample points on a uniform grid in $\mathcal{M}^{3/5}$ and $\eta$ while using the same sample strategy for the remaining five parameters. 

In the validation step we evaluate the model at randomly chosen parameter values. Parameter values at which the approximation error is greater than $10^{-6}$ are flagged. We combine these high-error points to the ones previously selected by the greedy algorithm; their union constitutes a new training set. Running the greedy algorithm on this new set produces an enriched basis with an improved error as judged by yet another series of validations. The validation$\rightarrow$enrichment$\rightarrow$validation$\rightarrow\dots$ iterations continue until the worst error
is below $10^{-6}$. This is somewhat similar-in-spirit to the sampling strategy of~\cite{Blackman2014} described above.

Due to the fact that the validation step is embarrassingly parallel over the random samples (as opposed to the greedy algorithm, which requires a modest amount of communication), we can easily handle a large number of random points. We typically consider roughly $10$ to $15$ million points per validation study.

\subsection{Frequency resolution of the training set}
\label{sec:TrainingSetQuadrature}

To capture the main waveform features the training set must be faithfully sampled in both parametric and physical dimensions. This must be balanced against the size of the training spaces in physical memory. For example, storing a training set with $(64^{2})\times(8^5)$ waveforms with a bandwidth of $\sim 4096\,$ Hz and a frequency resolution of $\Delta f = 1/64\,$ Hz would require around $500$ terabytes of memory. Our training set waveforms use an adaptive frequency sampling strategy, $\Delta f (f)$, which significantly reduces the greedy algorithm's memory footprint  to around 64GB. We only apply this adaptive sampling to cases D-F in Table~\ref{tab:basis_ranges} as the other cases' training sets fit comfortably into memory.

Our choice of frequency resolution $\Delta f (f)$ comes from determining the longest signal duration for a given mass and frequency band. These are found empirically, first by finding the duration of the lightest binary system in a set of frequency bands for each of the cases in Table~\ref{tab:basis_ranges}, and then rounding this up to the next-highest-power of two. The frequency resolution is taken to be the inverse of this duration. By selecting $\Delta f (f)$ in this way, we ensure that the waveforms are sampled above the Nyquist rate in each band. This can be applied across multiple bands (20Hz - 64Hz, 64Hz - 128Hz, etc.). This is a similar strategy to ``multibanding'' which has been useful in other contexts e.g., the gravitational-wave search pipeline of \cite{0004-637X-748-2-136} -- see Table~3 of Ref.~\cite{0004-637X-748-2-136}.

We stress that the adaptive frequency sampling described above is used for training set waveforms only. Once the greedy points are known, to collocate with the data on a set of equally spaced frequencies corresponding to the global Nyquist rate, we up-sample by direct evaluation of the waveform model. This does not cause a memory bottleneck, however, because the basis is significantly smaller than the training set. Refs.~\cite{PhysRevD.86.084046,antil2012two,PhysRevLett.114.071104} used a similar strategy whereby the frequency interval was split with a domain decomposition following the local Nyquist frequency and employing Gaussian quadratures in each subdomain. Additional validation is needed to check that up-sampling does not introduce an unacceptably large error, which we demonstrate in Sec.~\ref{sec:upsampling}. Refs.~\cite{antil2012two,PhysRevLett.114.071104} provide further discussions of subtleties related to up-sampled basis.

\subsection{The basis building pipeline}
\label{secd:RB}

The ROQ rule derived in Sec.~\ref{sec:roq_rule} requires a basis set for both the plus- and cross-polarizations, $\{B_j\}_{j=1}^{N_{\tt L}}$ in Eq.~(\ref{eq:EIM_lin}),
and another, different basis set for the three product combinations of these polarizations, $\{C_k\}_{k=1}^{N_{\tt Q}}$ in Eq.~(\ref{eq:EIM_quad}). We build these linear and quadratic parts of the ROQ hierarchically in steps.

\begin{figure}
\includegraphics[width=0.95\linewidth]{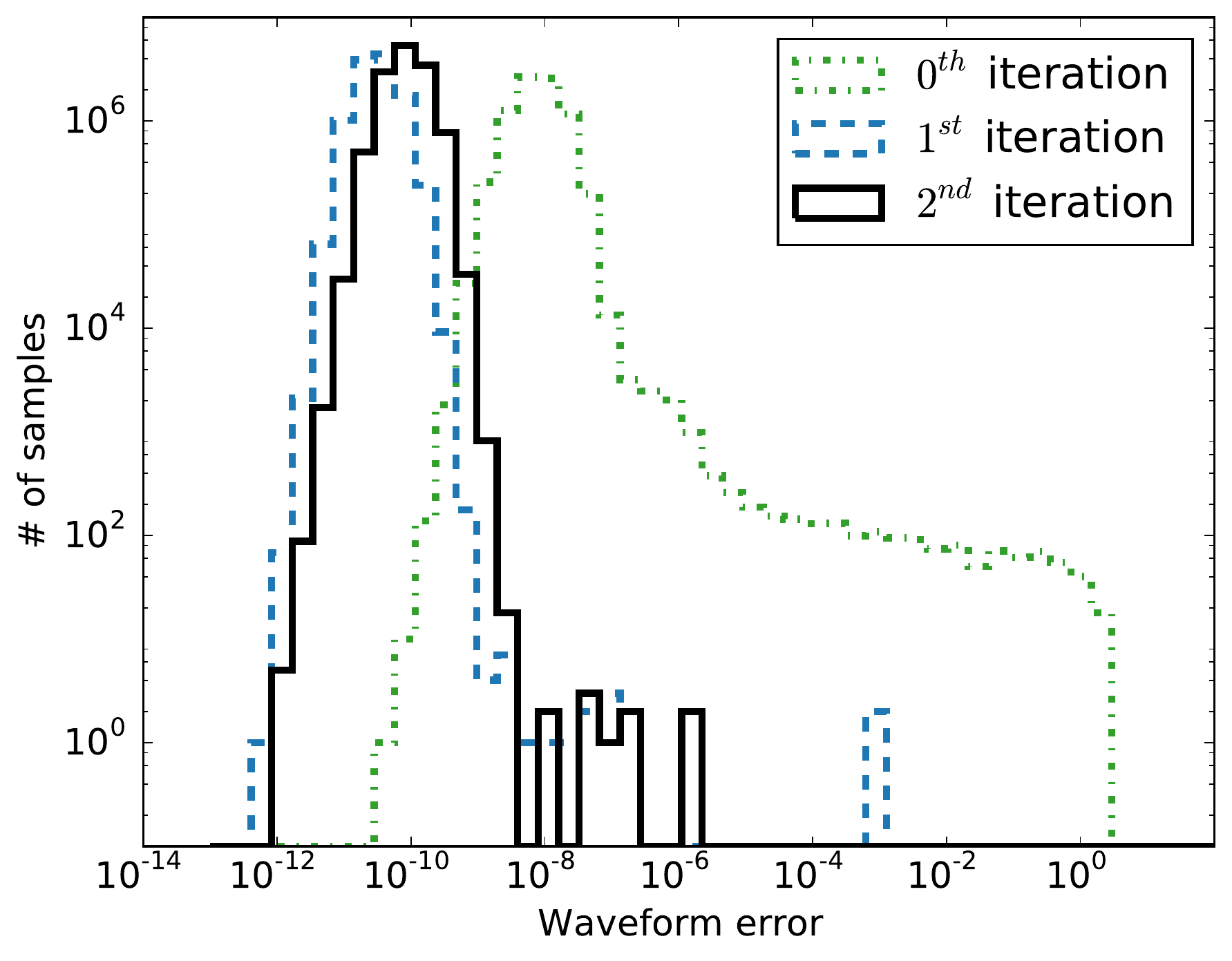}\\
\includegraphics[width=0.95\linewidth]{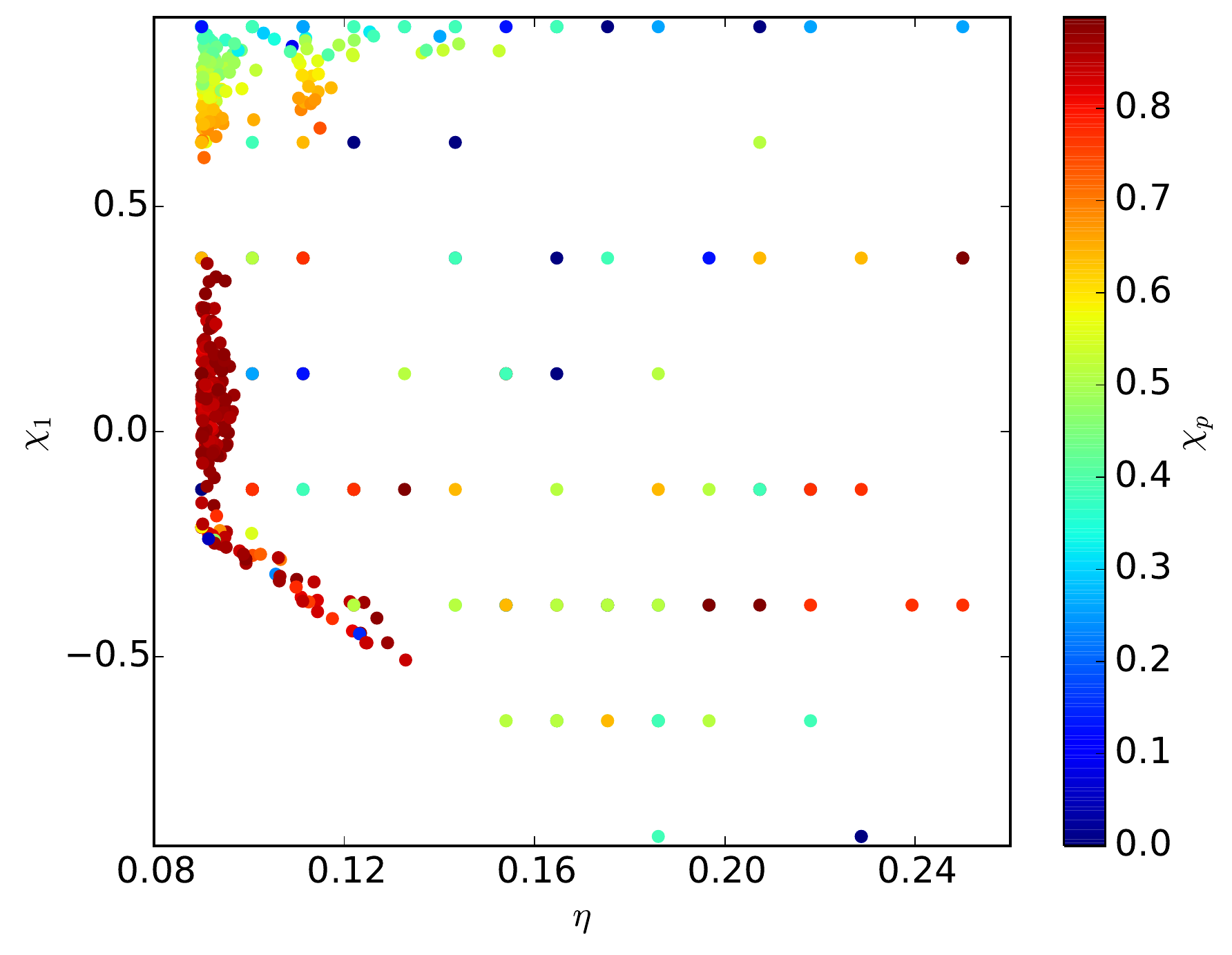}
\caption{An example of the basis enrichment strategy applied to case D from Table~\ref{tab:basis_ranges} (the plus- and cross-polarizations only). {\bf Top}: A sequence of histograms showing the distribution of the reduced basis approximation error for $\approx 10$ million out-of-sample model evaluations. Notice a continual lowering of the \textit{maximum} approximation error with each iteration. {\bf Bottom}: The final distribution of greedy points in a three-dimensional subspace. This set is a mixture of the initial structured grid used in the zeroth iteration and random points identified through the enrichment process.}
\label{fig:enrichment}
\end{figure}

We start by building a basis for the linear part of the ROQ. Empirically, we have found that an accurate basis trained exclusively for the plus-polarization continues to approximate the cross-polarization with good accuracy, and vice versa. 
Consequently, we populate a training set for the $\tilde{h}_{+}$ mode of the strain only. A greedy algorithm identifies a ``zeroth iteration" basis $\{B_j^0\}_{j=1}^{N_{\tt L}^0}$. We then perform a validation of this basis against both polarizations. Waveform errors greater than $\epsilon = 10^{-6}$ are used in a basis enrichment step described above. We iterate until an $\epsilon$-accurate basis is achieved (often $1$ or $2$ iterations suffice). Fig.~\ref{fig:enrichment} displays a sequence of error histograms (top panel) and the final distribution of greedy points (bottom panel) in a three-dimensional subspace.

Next, a basis for the quadratic part of the ROQ is built using previously computed information. To motivate our approach, notice that a tensor product of the linear basis is sufficient to describe $\tilde{h}^*_+\tilde{h}_+$, $\tilde{h}^*_\times \tilde{h}_\times$ and $\Re \left(\tilde{h}^*_+\tilde{h}_\times\right)$. Consequently, we take the greedy points which define the linear basis and form an ansatz training set consisting of $(\tilde{h}_+^{*} + \tilde{h}_\times^{*})(\tilde{h}_+ + \tilde{h}_\times)$. The quadratic basis $\{C_i(f)\}_{i=1}^{N_{\tt Q}}$ is built following the same iterative enrichment procedure used for the linear basis. A more direct (but more costly) two-step approach to treating these product terms is given in~\cite{antil2012two}. 

\subsection{Translating basis results to new regions of mass and frequency}
\label{sec:translate}

\begin{center}
\begin{figure}
\includegraphics[width=0.95\linewidth]{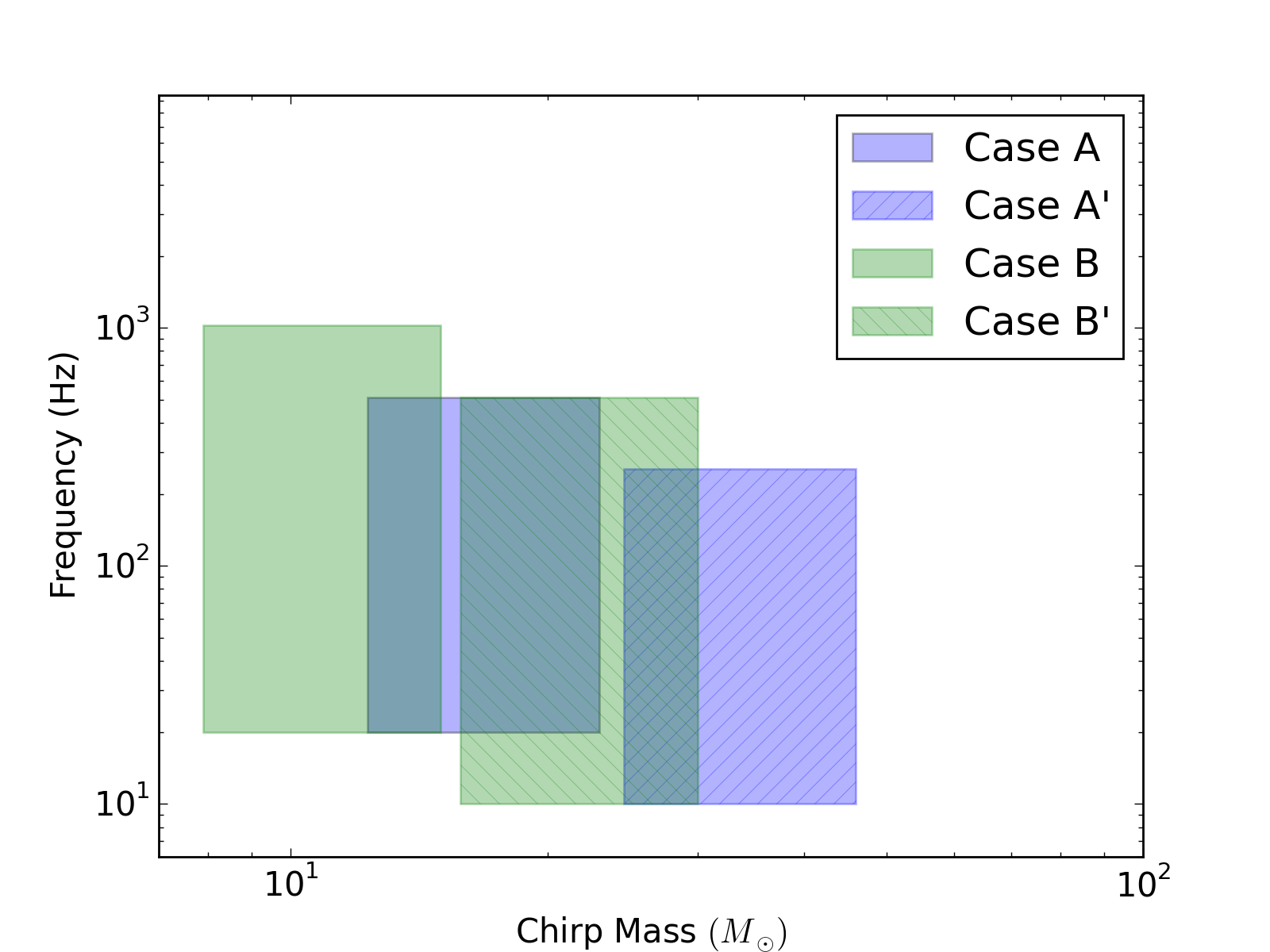}
\caption{Illustration of how to translate basis results to new regions of mass and frequency. By using the scaling described in the text, the basis for case A (blue unhatched region) maps on to case A$^{\prime}$ (blue hatched region). Similarly, the unhatched and hatched green regions respectively correspond to cases B and $B^{\prime}$ in Table~\ref{tab:basis_ranges}. These primed-regions form a starting point for building a $10$Hz basis without extra computational effort.}
\label{fig:scaling}
\end{figure}
\end{center}

By exploiting a mass-frequency mapping allowed by the Einstein equation in vacuum we can extend the basis' region of validity over an enlarged mass and frequency range which would otherwise require extra computational effort to build. Recall that a waveform described by a particular chirp mass $\mathcal{M}$ and low and high frequencies $f_{\text{low}}$ and $f_{\text{high}}$ can be transformed into a waveform described by a chirp mass of $\mathcal{M}^{\prime}=n\mathcal{M}$ and low and high frequencies of $f^{\prime}_{\text{low}} = f_{\text{low}}/n$ and $f^{\prime}_{\text{high}} = f_{\text{high}}/n$. As an example, consider the basis for case A (Table~\ref{tab:basis_ranges}) - which covers masses $12.3 \leq \mathcal{M}/M_{\odot} \leq 23$ and frequencies $20 \leq f/\text{Hz} \leq 512$ at a resolution of $\Delta f = 1/4$Hz - which can be mapped onto case A$^{\prime}$ with masses $24.6 \leq \mathcal{M}/M_{\odot} \leq 46$, frequencies $10 \leq f/\text{Hz} \leq 256$ with a frequency resolution of $\Delta f = 1/8$Hz by setting $n=2$. This procedure can be repeated to access higher masses and lower frequencies, or lower masses and higher frequencies. Table~\ref{tab:basis_ranges} summarizes one possible extension of a ROM/ROQ from $f_{\text{low}} = 20$Hz to $f_{\text{low}} = 10$ Hz. Fig.~\ref{fig:scaling} depicts the appearance of gaps in the translated ROM, the filling of which would require additional numerical work, although significantly less than had the $10$Hz-basis been built from scratch. This technique, which necessarily requires our basis have been built \textit{without} reference to any particular noise curve, has also been used in other ROMs~\cite{0264-9381-31-19-195010, Purrer:2015tud, PhysRevX.4.031006, PhysRevLett.115.121102}.

\section{Building and validating the empirical interpolant} 
\label{sec:build_and_valid}

In this section we numerically compute the requisite empirical interpolation representation~\eqref{eq:EIM} which, once the detector's data $d$ is known (cf.~Sec.~\ref{sec:PE}), will enable accelerated likelihood evaluations from Eq.~\eqref{eq:roq_simple}. Since the ROQ's error, $\left| \log\mathcal{L}(d|\vec{\Lambda}) - \log\mathcal{L}(d|\vec{\Lambda})_{\tt ROQ} \right|$, is controlled by the empirical interpolant's error~\cite{antil2012two}, we are especially interested in quantifying the latter error for any possible gravitational wave model evaluation. Given an approximation $\hat{a} \approx a$, which could stand for either the linear or quadratic parts, we report the error as the square of the un-weighted ($S_n = 1$) norm of $\hat{a} - a$, which is related to the un-weighted overlap, $(\hat{a}, a)$,  by 
\[
\| \hat{a} - a \|^2 = (\hat{a}- a,\hat{a}- a) \, ,
\]
where $\hat{a}$ and $a$ are normalized.
It is this ``white-noise" error which directly controls the ROQ's log-likelihood approximation error. The next section describes parameter estimation studies for which, clearly, $S_n \neq 1$.

\subsection{Linear parts}

Our first task is to build the basis, $\{B_j\}_{j=1}^{N_{\tt L}}$, and ROQ nodes, $\{F_j\}_{j=1}^{N_{\tt L}}$. These pieces are required to form the part of the ROQ rule~\eqref{eq:ROQ_part1} which is linear in $\tilde{h}$.

\begin{center}
\begin{figure}
\includegraphics[width=0.95\linewidth]{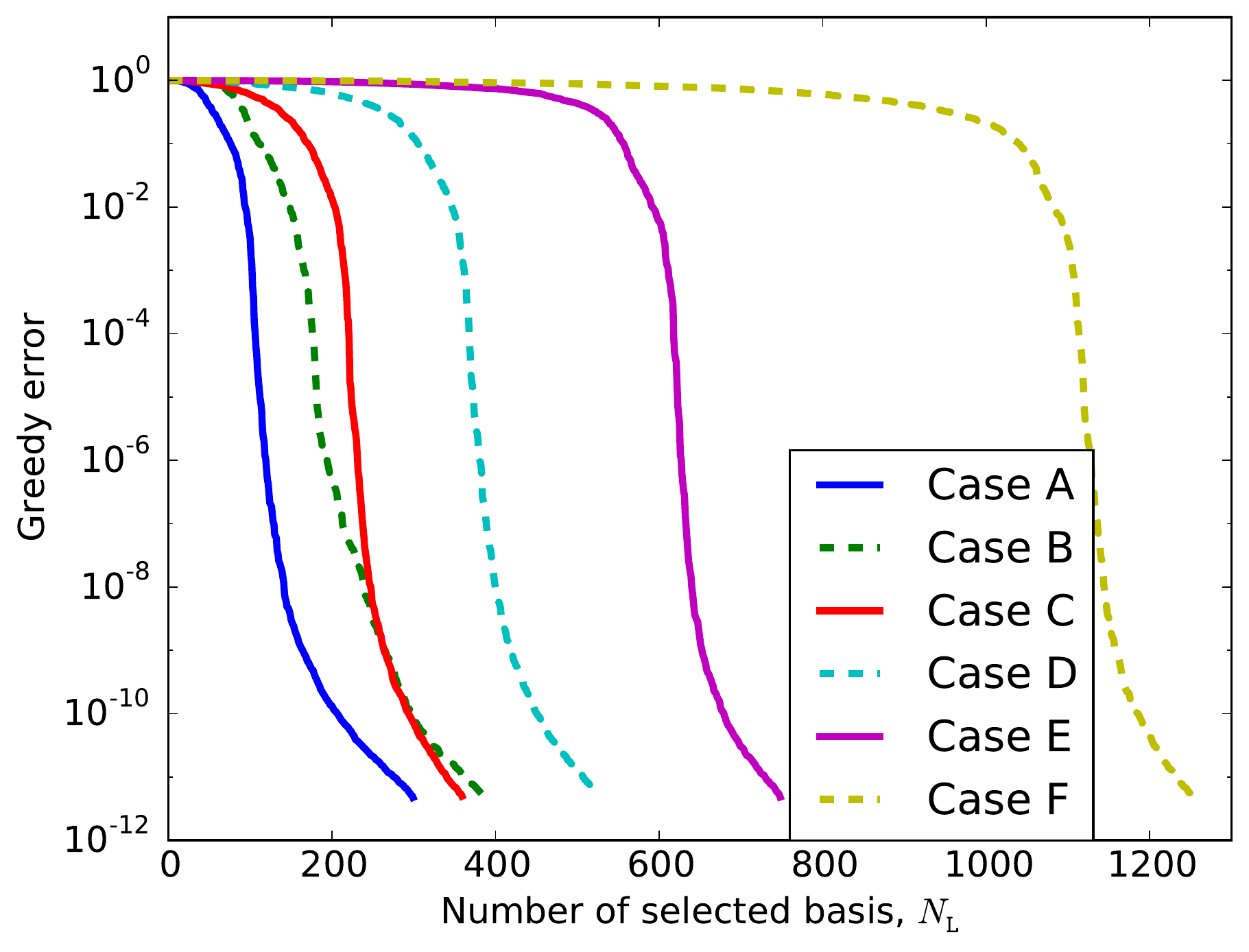}
\caption{Greedy error, defined as the maximum approximation error over the training set using the first $N_{\tt L}$ basis, versus basis number $N_{\tt L}$. The error profile is fairly similar across all cases. Since the ROQ speedup is (almost) proportional to $N_{\tt L}$, further speedup can be achieved at the expense of accuracy. Throughout the paper we select the first $N_{\tt L}$ basis satisfying a conservative $5\times10^{-12}$ greedy error tolerance.}
\label{fig:greedy_errs}
\end{figure}
\end{center}

To find the basis, we apply the greedy algorithm to training sets defined on each Case A-F from Table.~\ref{tab:basis_ranges}. As discussed in the previous section, our training set is iteratively enriched with random sampling. Fig.~\ref{fig:greedy_errs} reports the greedy algorithm's error profile when applied to the final (and hence most dense) training set iteration. Fig.~\ref{fig:greedy_errs} shows a similar behavior in all cases, namely, an initially slow fall-off in the representation error followed by an exponential decrease. This by-now common feature has been seen across different waveform models using different dimensional reduction algorithms~\cite{PhysRevD.86.084046, Caudill:2011kv, Blackman2014, Purrer:2015tud, PhysRevLett.114.071104, PhysRevLett.114.071104, PhysRevLett.115.121102, 0264-9381-31-19-195010, PhysRevX.4.031006, cannon2010singular}. Fig.~\ref{fig:greedy_errs} also shows that the number of reduced basis waveforms needed to approximate intervals describing successively smaller chirp mass values increases. A notable exception, however, is case B which for some error thresholds is actually larger than Case C. One possible explanation is that the iteratively enriched basis is sub-optimal as compared to a hypothetical basis built from an arbitrarily dense training set. Nevertheless, even these sub-optimal basis provide excellent performance gains while being less demanding to compute. Table~\ref{tab:basis_ranges} summarizes the resulting linear bases corresponding to a greedy error of $5\times 10^{-12}$.

\begin{center}
\begin{figure}
\includegraphics[width=0.85\linewidth]{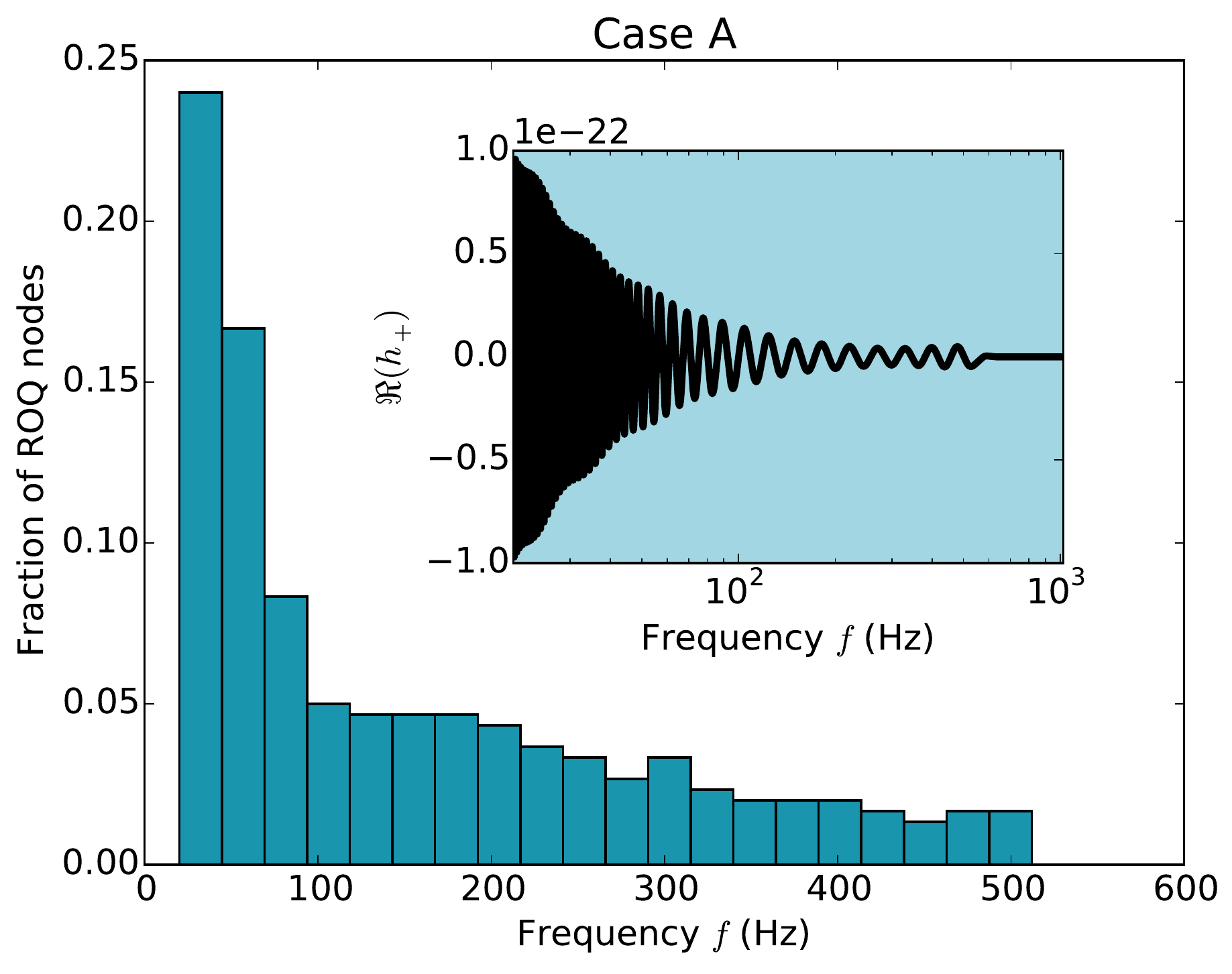} \\
\includegraphics[width=0.85\linewidth]{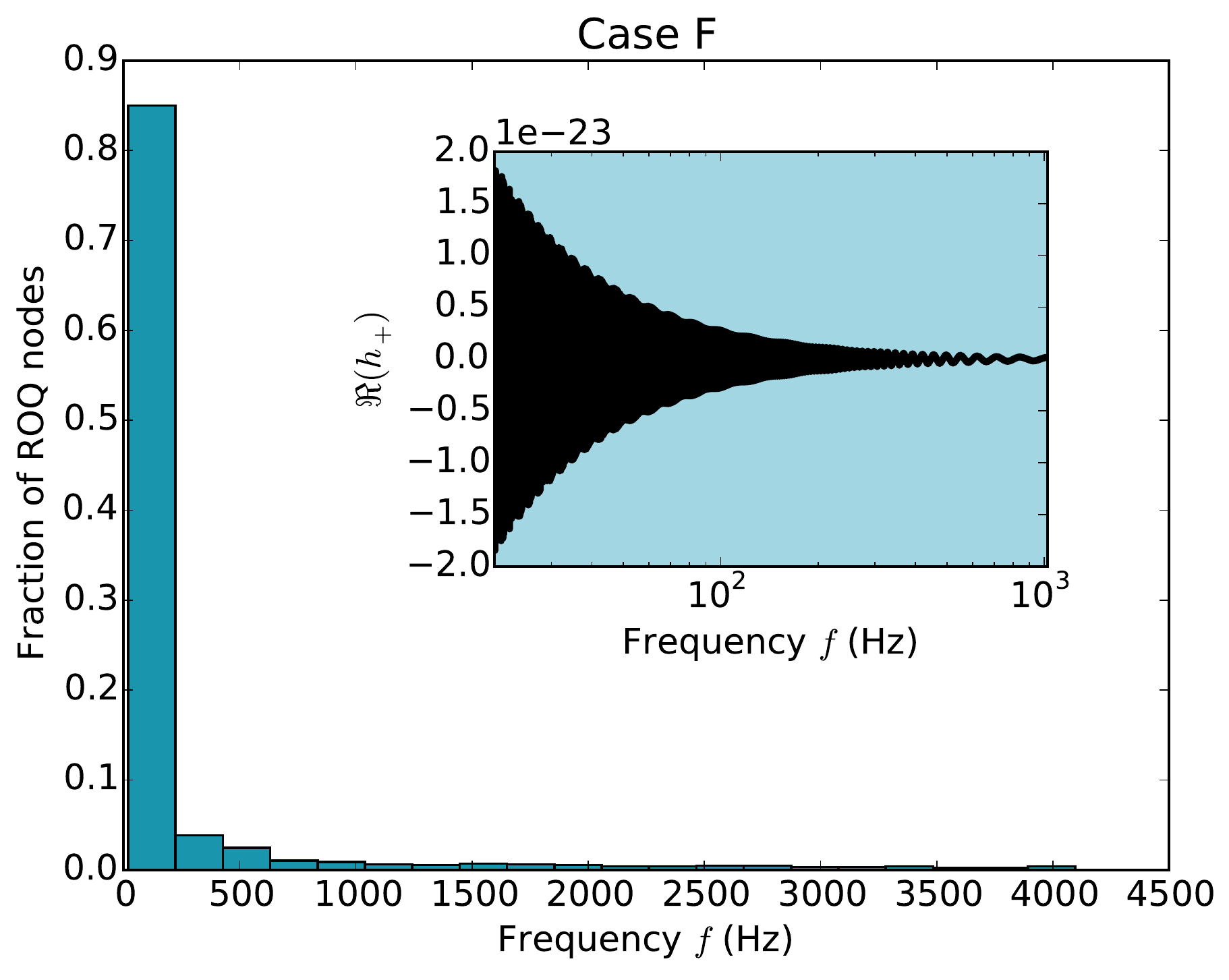} 
\caption{Histogram of selected ROQ nodes and a representative waveform for case A (top) and case F (bottom). Evidently the selected frequency points cluster at small values. This is intuitively expected because lower frequency intervals contain a greater number of waveform cycles, a feature which is automatically detected by the empirical interpolation method.
Histograms of those cases not shown are qualitatively similar, being a mixture of these two boundary cases.}
\label{fig:roq_histogram}
\end{figure}
\end{center}

To find the ROQ nodes, we apply the empirical interpolation method for each case defined in Table.~\ref{tab:basis_ranges}. As input to the algorithm we provide the reduced basis vectors and the corresponding set of frequency points. Figure.~\ref{fig:roq_histogram} depicts the distribution of selected ROQ nodes for the two most extreme cases A (top) and F (bottom). Notice the that EI method preferentially selects points at lower frequencies, which matches our expectation that the information carried by these waves is encoded in the cycles which ``pile up" at lower frequencies.

\begin{center}
\begin{figure*}
\includegraphics[width=0.95\linewidth]{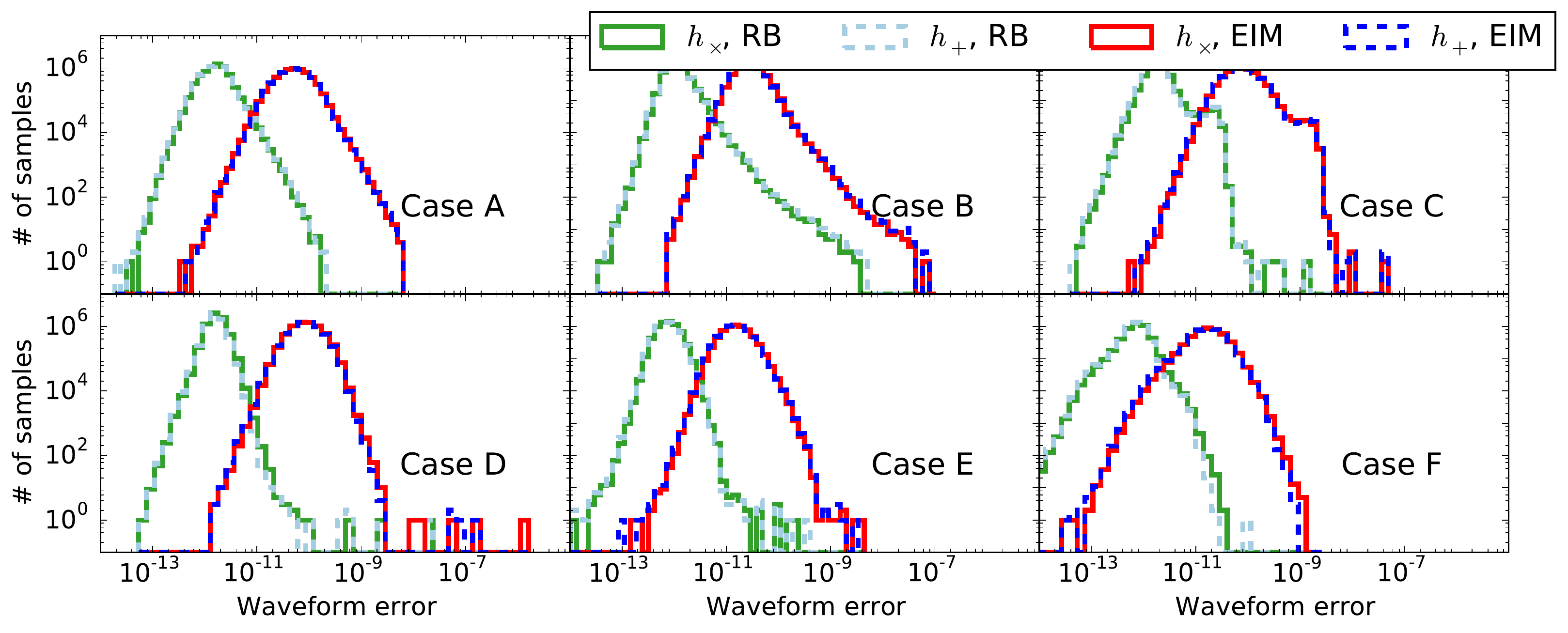}
\caption{Projection (RB) and empirical interpolation (EIM) errors (generically x-axis labeled as ``Waveform error") for $\approx 15$ million randomly drawn waveforms. Each subfigure reports on the errors for an approximation defined by the six cases listed in Table ~\ref{tab:basis_ranges}. The validations
are performed using the same adaptive frequency sampling strategy as was used to find the basis (cf.~Sec.~\ref{sec:TrainingSetQuadrature}).}
\label{fig:ValidationStudy1}
\end{figure*}
\end{center}

Fig.~\ref{fig:ValidationStudy1} reports the out-of-sample validation study, which uses $\approx 15$ million random waveform evaluations not in the original training set. The errors $\epsilon_{\times} = ( \hat{h}_\times - h_\times, \hat{h}_\times - h_\times)$ and $\epsilon_+ = ( \hat{h}_+ - h_+, \hat{h}_+ - h_+)$ are found to be small in all cases. Thanks to the frequency-independence of the antenna patterns, one can directly relate these errors to the error in the linear part of the ROQ rule~\eqref{eq:ROQ_part1}
\begin{align*}
&\left| ( d, h) - ( d, h)_{\tt ROQ}  \right|  
\leq {\cal C}_1 \left| F_+ \right| \epsilon_+ + {\cal C}_2 \left| F_{\times}\right| \epsilon_{\times}  \,, \nonumber
\end{align*}
without any extra numerical work. Importantly, this avoids the computation of errors over an enlarged parameter space including $\text{ra}$, $\text{dec}$ and $\psi$. The constants ${\cal C}_1$ and ${\cal C}_2$ are computable. Finally, Fig.~\ref{fig:ValidationStudy1} demonstrates that we incur a penalty factor of $\approx 100$ when approximating by an empirical interpolant as opposed to orthogonal projection onto the basis which is guaranteed to yield the smallest possible error. We do not know ahead of time what this penalty factor might be; this further motivates our choice of working to small $5\times 10^{-12}$ accuracies in the basis building step.

\subsection{Quadratic parts}

\begin{center}
\begin{figure*}
\includegraphics[width=0.95\linewidth]{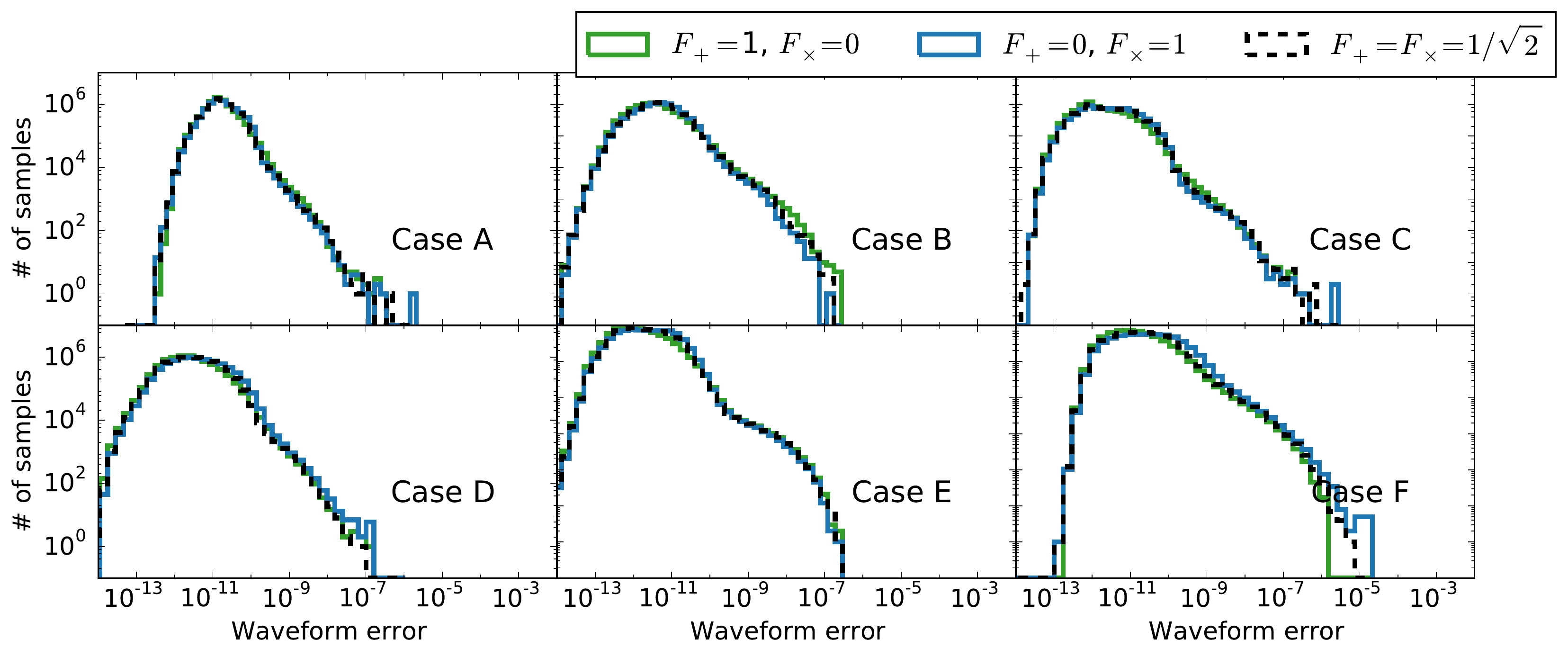}
\caption{Empirical interpolation errors (generically x-axis labeled as ``Waveform error") for $\approx 15$ million randomly drawn waveforms. Each subfigure reports on the errors for an approximation defined by the six cases listed in Table ~\ref{tab:basis_ranges}. The validations are performed using an adaptive frequency sampling strategy and for three representative antenna pattern configurations.}
\label{fig:ValidationStudy2}
\end{figure*}
\end{center}

Our next task is to build the basis, $\{C_j\}_{j=1}^{N_{\tt Q}}$, and ROQ nodes, $\{\mathcal{F}_j\}_{j=1}^{N_{\tt Q}}$. These pieces are required to form the part of the ROQ rule~\eqref{eq:ROQ_part2} which is quadratic in $\tilde{h}$. The steps are essentially the same as in the linear case just described. Table~\ref{tab:basis_ranges} summarizes the resulting quadratic bases corresponding to a greedy error of $5\times 10^{-12}$.

We now skip directly to the approximation errors, quantified by yet another out-of-sample validation study. As for the linear case, we would again like to relate the ROQ error to the errors due to approximation of each quadratic polarization parts. Due to the differences in sizes of each quadratic piece, computing relative errors are uninformative in this case~\footnote{Since $\|\Re \left(\tilde{h}^*_+\tilde{h}_\times\right) \| \ll \| \tilde{h}^*_+\tilde{h}_+ \|$ in the non-precessing limit, the relative approximation error of 
$\Re \left(\tilde{h}^*_+\tilde{h}_\times\right)$ may be large but insignificant insofar as likelihood accuracy is concerned.}. We instead compute the error from the approximation of $\left(F_+ \tilde{h}_+ + F_\times \tilde{h}_\times\right) \left(F_+ \tilde{h}_+^* + F_\times \tilde{h}_\times^*\right)$ by its empirical interpolant for three representative cases. The results are shown in Fig.~\ref{fig:ValidationStudy2}.

\subsection{Upsampling}
\label{sec:upsampling}

\begin{center}
\begin{figure}
\includegraphics[width=0.95\linewidth]{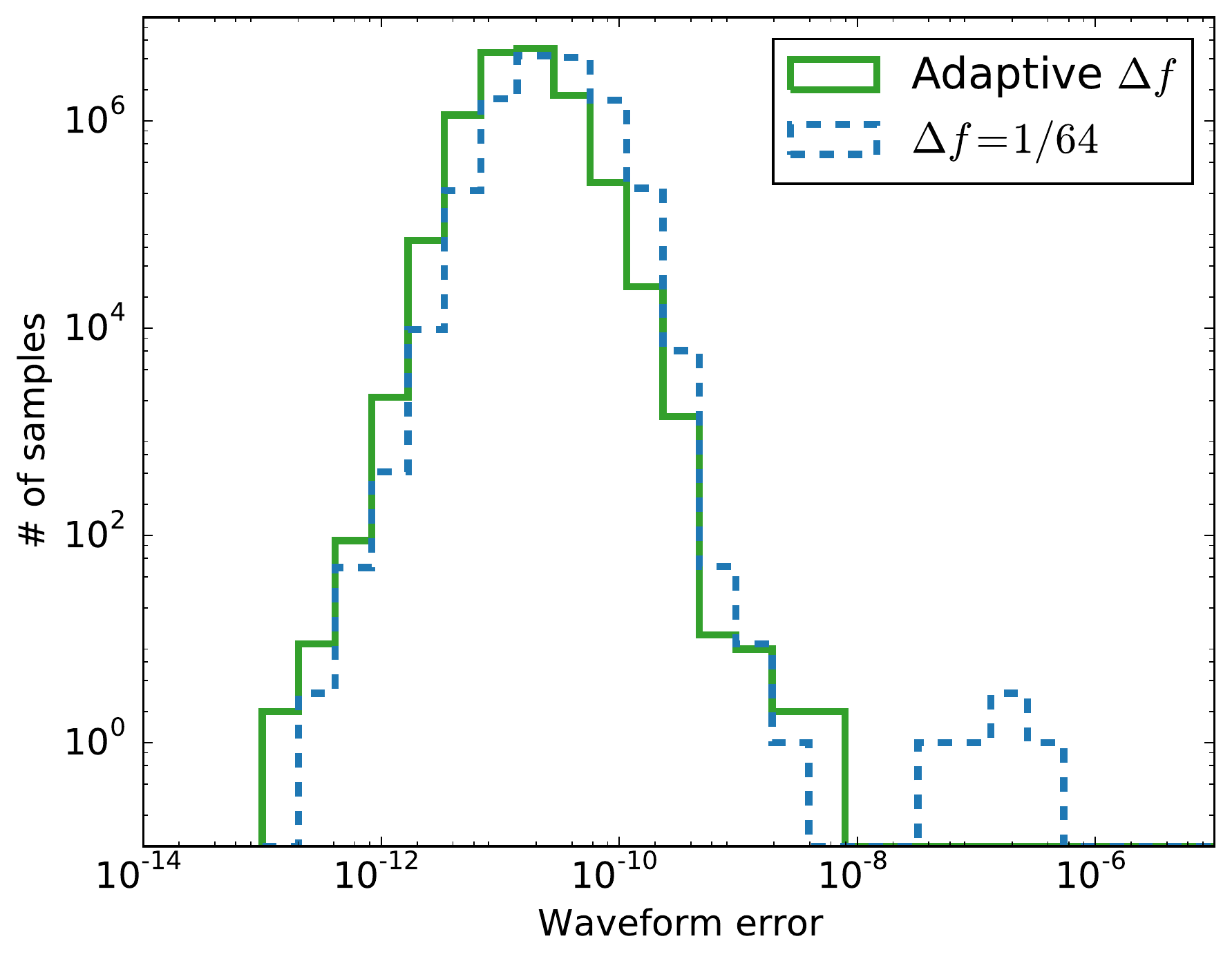}
\caption{Empirical interpolant approximation errors (the plus- and cross-polarizations only) when using an adaptive (solid green line) and uniform (dashed blue line) frequency sampling. The adaptive sampling is used during the ROQ building procedure. To compute log-likelihoods with our ROQ rule we upsample to a uniform frequency grid, and so this error (which constitutes the last in a series of approximations of the underlying model) is the most relevant for ROQ-accelerated inference studies. Results are shown for Case E only; other cases are qualitatively similar. Maximum upsampled EIM errors of $6\times10^{-9}$, $1\times10^{-7}$, $1\times10^{-5}$, $7\times10^{-8}$, $4\times10^{-7}$ and $1\times10^{-9}$ were computed for Cases A-F, respectively.}
\label{fig:upsample_err}
\end{figure}
\end{center}

As discussed in Sec.~\ref{sec:TrainingSetQuadrature}, in order to reduce the greedy algorithm's memory footprint to manageable sizes we use an adaptive frequency sampling strategy. Yet to compute the ROQ weights~(\ref{eq:ROQ_part1}a, \ref{eq:ROQ_part2}a) the basis must be known at the same frequency values recorded by the detector. To collocate with the data on a set of equally spaced frequencies corresponding to the global Nyquist rate, we up-sample by direct evaluation of the waveform model at the greedy points and reorthogonalize the basis. Fig.~\ref{fig:upsample_err} reports the additional error due to up-sampling. That the errors remain similarly small is evidence that our training set waveforms are well-resolved by the adaptive frequency grid.

\section{Parameter Estimation} 
\label{sec:PE}
\subsection{Accuracy comparisons}

To determine how the empirical interpolation errors (as summarized in Figs.~\ref{fig:ValidationStudy1}, \ref{fig:ValidationStudy2}, \ref{fig:upsample_err}) affect parameter estimation, we present a comparison between the recovered posterior PDFs using both the \textit{Full} and the ROQ likelihood functions evaluated with  \texttt{LALInferenceNest} \cite{PhysRevD.91.042003}, which is one of the stochastic samplers available with the \texttt{LALInference} library \cite{LAL}. A simulated binary black hole signal represented by IMRPhenomPv2 and drawn from the parameter space defined by case A in Table~\ref{tab:basis_ranges} was injected coherently in the two LIGO detectors. To represent the non-stationarity of the detector noise the injection was made into real data from the sixth LIGO science run \cite{0264-9381-32-11-115012}, recoloured to reflect the expected early aLIGO sensitivity (cf. Ref.~\cite{s6_public_noise_curves,Berry:2014jja} which used the same data for studying simulated binary neutron star detections).

\begin{center}
\begin{figure*}
\includegraphics[width=0.48\linewidth]{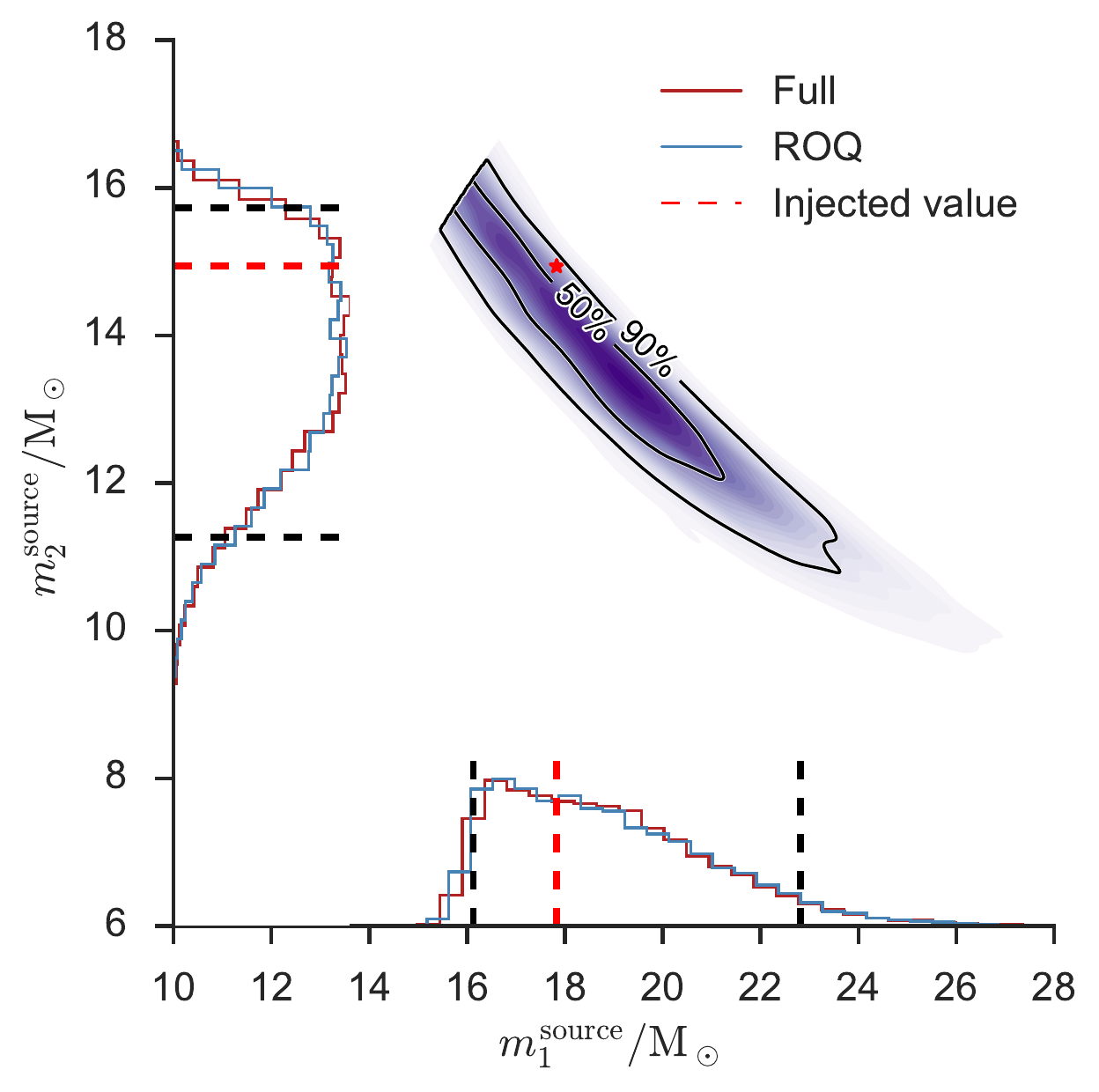}
\includegraphics[width=0.48\linewidth]{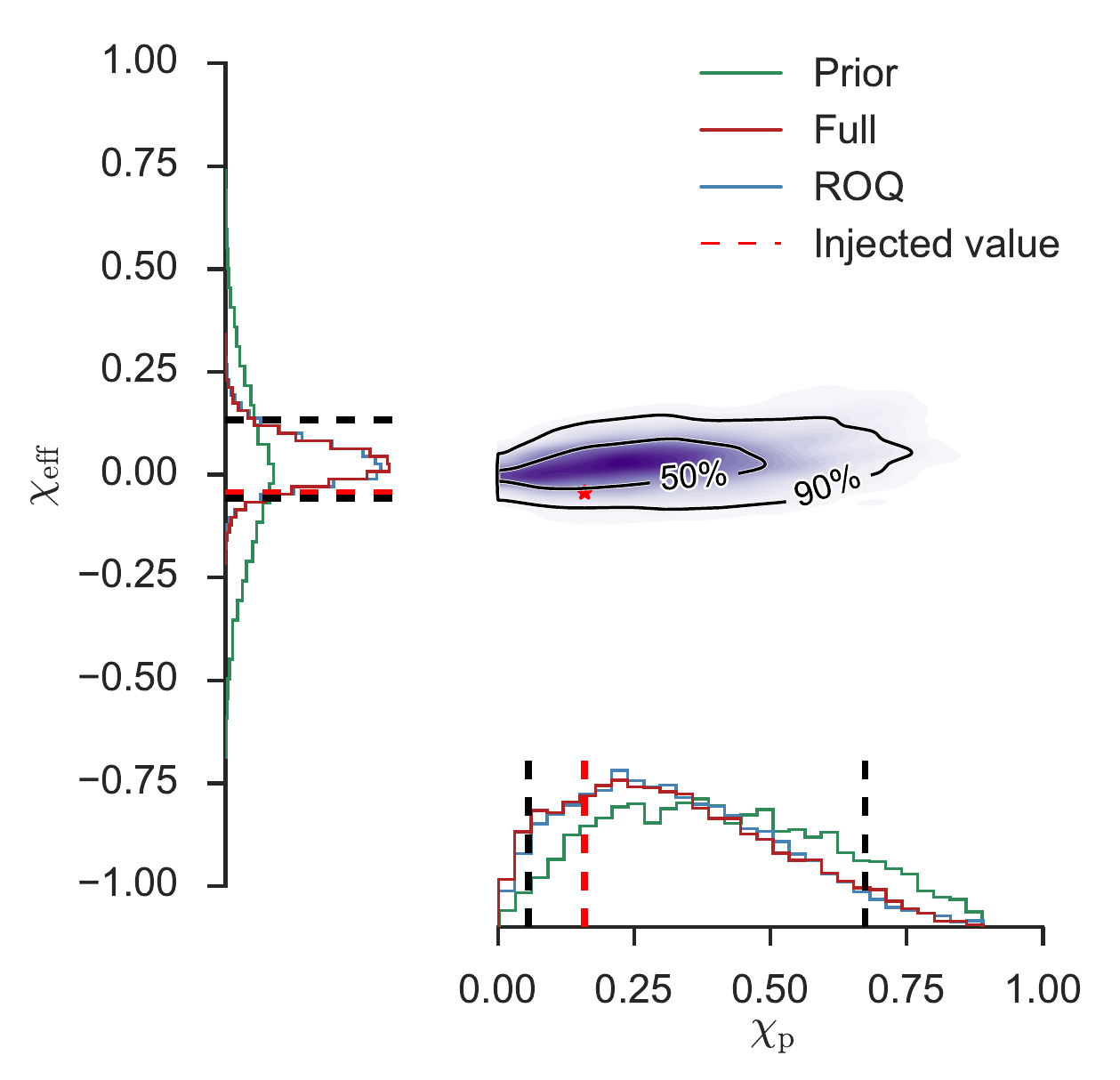}
\caption{Comparing the recovered posterior PDFs for the black hole masses measured in their source frame (Left) and the two dominant spin parameters (Right), c.f. \cite{TheLIGOScientific:2016wfe}. The parameter $\chi_{\text{eff}} = (m_1\chi_1 + m_2\chi_2)/(m_1 + m_2)$ is known as the ``effective'' spin and is a mass-weighted combination of the two spin components parallel to the orbital angular momentum. The dashed black lines mark the $90\%$ credible interval for both the ROQ and Full likelihoods, which are the same within the statistical sampling uncertainty of $\sim 1\%$. The posteriors do not peak exactly at the injected values due to the presence of detector noise.}
\label{fig:PEaccuracy}
\end{figure*}
\end{center}

Under the assumption that the ROQ is an approximation of the Full likelihood, 
the two methods are required to be statistically indistinguishable in order for the the ROQ to qualify as a valid substitute to the Full likelihood function for parameter estimation.
As is shown in Fig.~\ref{fig:PEaccuracy} the Full and the ROQ methods recover posterior PDFs that are almost visually identical. We quantify the difference between the two sets of posterior PDFs by computing the KL-divergence \cite{kullback1951}
\beq
D_{KL}(P|Q) = \sum_i P_i \log\left(\frac{P_i}{Q_i}\right) \,,
\eeq
for all of the one-dimensional PDFs produced by the Full and ROQ analyses, including but not limited to the parameters shown in Fig.~\ref{fig:PEaccuracy}. The KL-divergence quantifies the relative entropy, in units of bits, between the probability distributions $P$ and $Q$, or equivalently the amount of information lost when using $Q$ as an approximation to $P$. For $(P, Q)=$ (Full, ROQ) the minimum, median and maximum $D_{KL}$ are $(0.0020, 0.0057, 0.0141)$ bits respectively. This can be compared to the set of $D_{KL}$ for $(P, Q)=$ (Full, prior) of $(0.016, 0.33, \infty)$ bits, which reflects the information gain contained in the likelihood on its own. 

A comprehensive study of the parameter estimation capabilities using ROQs will be presented in \cite{ROQ_PE_followup_letter}.

\subsection{Performance Benchmarks}
\label{sec:performance}

\begin{figure}
\includegraphics[width=1.0\linewidth]{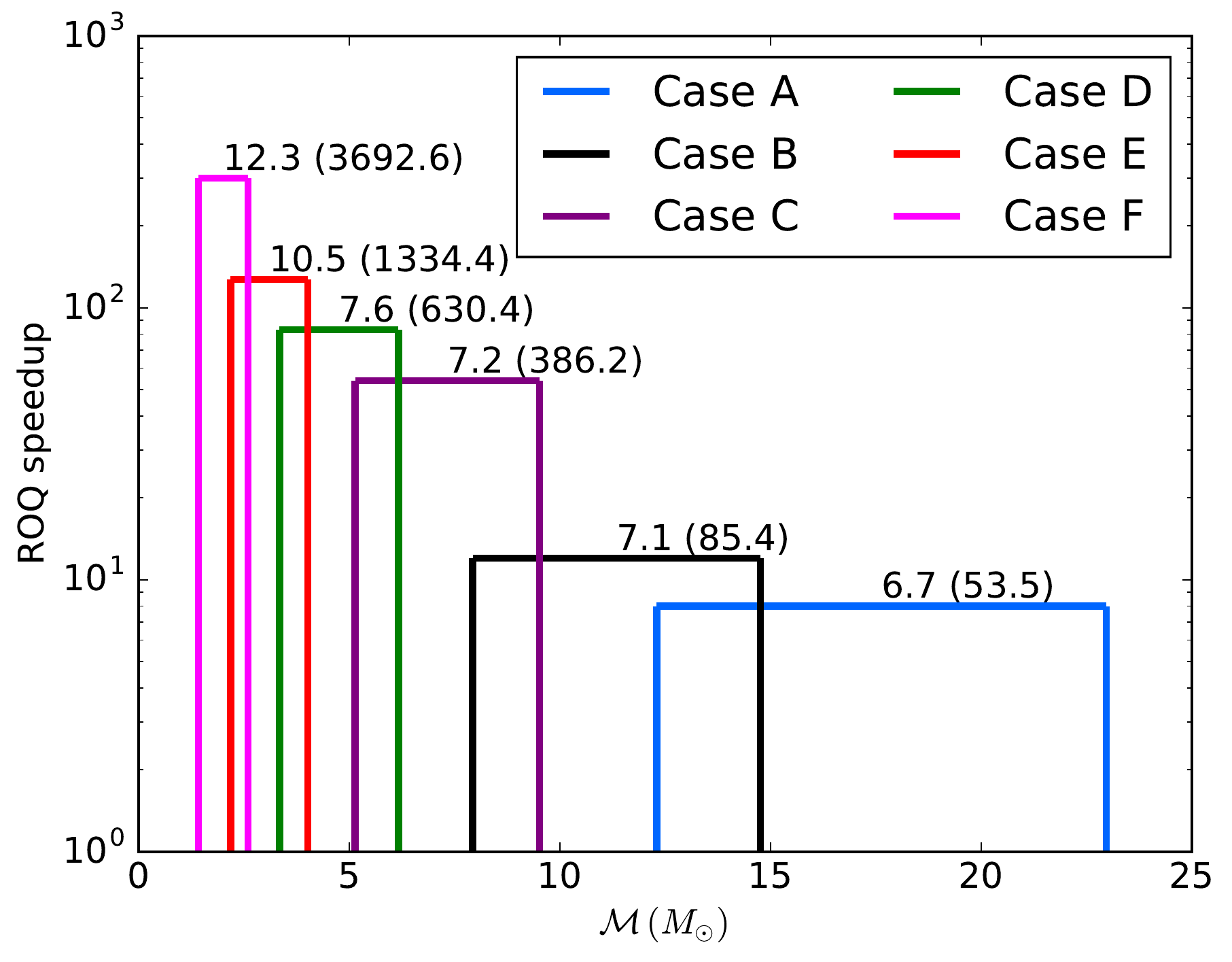}\\
\caption{Theoretical parameter estimation speedup (using the ROQ)
for cases A-F in Table~\ref{tab:basis_ranges}. The speedup is calculated from the ratio $L/(N_{\texttt{L}} + N_{\texttt{Q}})$, where $L = (f_{\text{max}} - f_{\text{min}})/\Delta f$ is the number of quadrature points in the Full likelihood, $N_{\texttt{L}}$ is the size of the linear basis and $N_{\texttt{Q}}$ is the size of the quadratic basis. The sum $N_{\texttt{L}} + N_{\texttt{Q}}$ is the number of points in the
ROQ likelihood~\eqref{eq:roq_simple}. The plot is annotated with the time (in hours) to compute $2\times 10^{7}$ ROQ (Full) likelihood evaluations, roughly the number of evaluations required for a typical PE analysis \cite{PhysRevD.91.042003}. Our tests were performed using an Intel Xeon CPU with a 2.70GHz clock speed.}
\label{fig:speedup_runtimes}
\end{figure}

Having established the equivalence of the results for the Full and ROQ likelihoods, we now consider the performance gains afforded by the ROQ rule. In Fig.~\ref{fig:speedup_runtimes} we show the expected likelihood speedup ratio $L/N$. Here $L$ is the number of operations in the non-ROQ likelihood and $N = N_{\texttt{L}} + N_{\texttt{Q}}$ is the number of operations in the ROQ likelihood~\eqref{eq:roq_simple}. The speedup is seen to be as large as $\approx 300$ for low mass systems. Assuming the entirety of the PE cost is in the form of waveform/likelihood evaluations, which scales linearly with $L$ (Full) or $N$ (ROQ), the ratio $L/N$ provides the theoretical performance improvement for any hypothetical PE study.

We estimate the run time of parameter estimation studies by $(i)$ computing the waveform at the empirical interpolation nodes for the linear and quadratic pieces of the ROQ, and $(ii)$ subsequently computing $2\times10^{7}$ evaluations of the ROQ-likelihood~\eqref{eq:roq_simple} for random-valued integration weights, which is a reasonable number of MCMC samples needed to produce a few thousand statistically independent samples using the \texttt{LALInference} code \cite{PhysRevD.91.042003}. These timing results are also summarized in Fig.~\ref{fig:speedup_runtimes}. We find that by using the ROQ, and assuming that the bulk of the cost of parameter estimation is in computing waveforms and overlap integrals, then the run time of PE codes should be between around $6$ hours (for analyses that restrict themselves to chirp mass bins as in case A of Table~\ref{tab:basis_ranges}) to around $12$ hours (for analyses that restrict themselves to chirp mass bins as in case F of Table~\ref{tab:basis_ranges}). Our tests were performed using a single core on an Intel Xeon CPU with a 2.70GHz clock speed. The test used a stand-alone python script calling the \texttt{LALSimulation} library through its SWIG interface.

These timing experiments obviously depend strongly on the effort of (hardware-specific) optimization or parallelization schemes, such as offloading work to MIC processors~\cite{MIC,ICC,MKL}, which we have not explored. Nevertheless, the quoted speedup numbers are \textit{independent} of these details.

Finally, we note that there is a once-per-analysis ``start up'' cost of computing the set of ROQ weights~(\ref{eq:ROQ_part1}b,\ref{eq:ROQ_part2}b). This cost, which amounts to $\mathcal{O}\left(10^4\right)$ overlaps~\eqref{e:inner} and parallelizes trivially, is negligible compared to a full inference simulation. As a representative example, we computed $10\,,000$ sets of ROQ weights for a typical time-window of $0.2$s centered on the trigger-time, each associated with a unique value of the coalescence-time $t_c$ within this window. Computing weights for $10\,,000$ values of $t_c$ corresponds to sampling the $t_c$ at a constant rate $\Delta t_c = 0.2/10^5 = 2\times 10^{-6}$, which is around a thousand times smaller than the typical measurement uncertainty in $t_c$ \cite{PhysRevLett.114.071104}. We find that the time to compute the ROQ weights is on the order of a few minutes for all cases in Table~\ref{tab:basis_ranges}, which is much smaller than both the estimated ROQ and Full inference run times. 

\section{Conclusion}
We have presented a method for building reduced order models and quadrature rules of precessing, inspiral-merger-ringdown gravitational waveforms designed specifically to improve the efficiency of astrophysical inference. Our method, which is generic, was applied to the waveform family known as IMRPhenomPv2. We find that by using an IMRPhenomPv2-specific reduced order quadrature rule, parameter estimation studies can be sped up by between factors of $4$ (for binary black holes) to $300$ (for binary neutron stars) in analyses starting from a low-frequency cutoff of 20Hz; see Fig.~\ref{fig:speedup_runtimes}. Crucially, this performance-boosting technique does not sacrifice the accuracy of parameter estimates as shown in Fig.~\ref{fig:PEaccuracy} and discussed in Sec.~\ref{sec:PE}. We stress that nearly-indistinguishable PE results are a consequence of the high accuracy ROM built in Sec.~\ref{sec:build_and_valid}. Below we discuss extensions to the work presented here.

\textit{Larger parameter regions}. The method presented here is generic and capable of handling large parameter domains. Recently, the non-precessing IMRPhenomD model~\cite{Husa:2015iqa, Khan:2015jqa} underlying IMRPhenomPv2 has been calibrated up to mass ratios of $q=18$ and aligned spins of $\sim 0.85$ ($0.98$ at equal-mass). We hope to explore the application of our methods to these extremal values of the model, which might require more sophisticated parameter sampling and domain decomposition strategies.

\textit{Other waveform families}. Some waveform families are described by costly differential equations. These could be effective-one-body models~\cite{Taracchini:2012ig,Pan:2013rra,buonanno1999effective,damour2008faithful}, PN models~\cite{blanchet2006gravitational,PhysRevD.80.084043} or the Einstein equations. While in principle our techniques can be applied to these models to construct the reduced basis and empirical interpolation nodes, it is not clear how to directly evaluate the waveform model \textit{at} the empirical interpolation nodes so that the ROQ can actually be used. As long as the ROM depends linearly on its basis, the surrogate modeling tools of 
Refs.~\cite{PhysRevD.87.122002,PhysRevD.87.044008,PhysRevD.85.081504,0264-9381-31-19-195010,Purrer:2015tud} may be applicable. Common to these techniques is the construction of a closed-form expression capturing the parametric behavior of well-chosen waveform data, such as the amplitude and phase values at specially selected times or frequencies. Consequently, the cost of evaluating a surrogate model will necessarily grow with parametric dimensionality. The efficiencies of these models for precessing systems remains an open question (none have been built to date). Currently, then, closed-form phenomenological waveform families offer the best trade off for achieving rapid and accurate parameter estimation with an ROQ. We believe ROQs to be especially useful for long waveforms dominated by many inspiral cycles, where approximate methods are expected to be accurate and ROQ speedups are at their largest.

\section{Acknowledgements}

We thank Harbir Antil, Jonathan Blackman, Thomas Dent, Chad Galley, Mark Hannam, Tom Loredo, Saul Teukolsky, Manuel Tiglio and Alan Weinstein for many useful discussions and encouragement
throughout this project, and Jonathan Blackman, Mike Boyle, Sascha Husa, and Alejandro Boh{\'e} for help towards explaining features described in Appendix~\ref{app:features}. We would also like to thank our LIGO Presentation and Publication reviewer for their clear and detailed feedback on this manuscript.

LIGO was constructed by the California Institute
of Technology and Massachusetts Institute of
Technology with funding from the National Science
Foundation and operates under cooperative agreement
PHY-0757058. MP was supported by STFC grant ST/I001085/1 and the Max Planck Gesellschaft.
SF was supported in part by NSF grants PHY-1306125 and AST-1333129 to Cornell University
and the Sherman Fairchild Foundation.
PS was supported by the Sherman Fairchild Foundation and NSF Grants No.
PHY-1404569 and PHY-1151197 at Caltech.
Some of the computations were carried
out using the high performance computing
resources provided by Louisiana State University (http://www.hpc.lsu.edu),
the Extreme Science and Engineering Discovery Environment (XSEDE) \cite{10.1109/MCSE.2014.80},
and the Zwicky cluster at Caltech, which is supported by the Sherman Fairchild Foundation
and by NSF award PHY-0960291.
We are grateful for computational resources provided by Cardiff University, and
funded by an STFC grant supporting UK Involvement in the Operation of Advanced
LIGO.

This paper carries LIGO Document Number
P1600096.
\appendix

\section{Greedy feature detector} \label{app:features}

Here we describe a novel use for the greedy algorithm which we believe might help waveform developers identify abrupt changes in behavior or discontinuities in waveform models.

One of the key criteria for the reduced basis method to deliver a basis that exhibits exponentially fast error convergence is that the model space varies smoothly with respect to parameter variations. When this criterion is not met, and the model space exhibits abrupt or discontinuous behavior, we typically find that the greedy algorithm selects basis elements from regions in parameter space where the non-smoothness occurs.  

We can use this to our advantage: by simply inspecting the location of points selected by the greedy algorithm and monitoring for high density clusters. This technique was previously used to find a problem in SEOBNRv1~\cite{Taracchini:2012ig} (see Fig. 15 in Ref.~\cite{0264-9381-31-19-195010}).

\begin{figure}[!t]
  \centering
    \includegraphics[width=\linewidth]{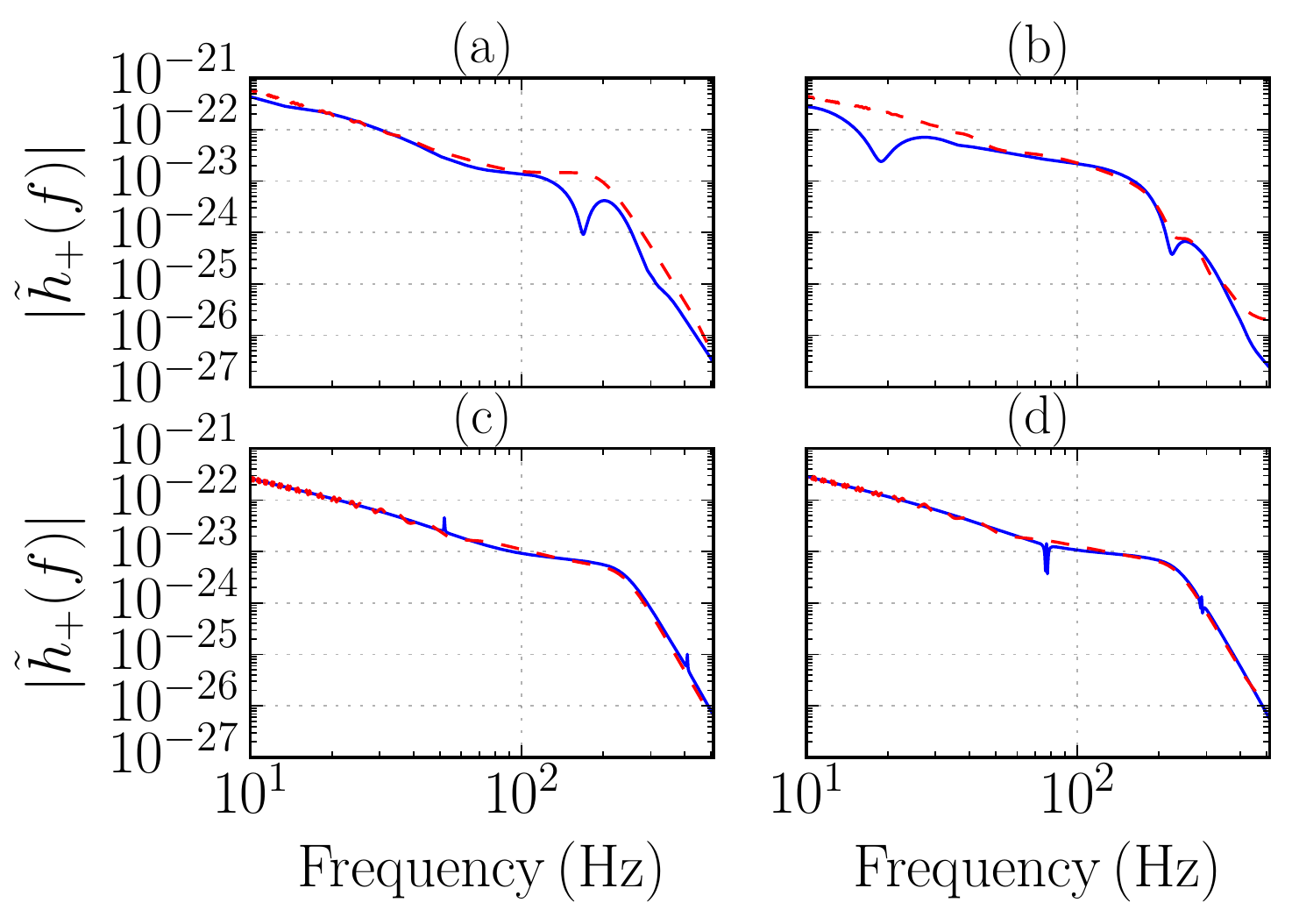}
    \caption{Amplitudes of $\tilde{h}_{+}(f)$ for selected points in the $\chi_\mathrm{p} \approx 0$ cluster shown in Fig.~\ref{fig:clusters}. The parameter values for configurations (a) -- (d) are given in Table~\ref{table:features-configurations}. Abrupt, sharp features are clearly visible in the IMRPhenomPv2 amplitudes (blue solid lines), but are absent in the SEOBNRv3 amplitudes (red dashed lines). These features are difficult to capture with the reduced basis method without sacrificing the sparsity and/or accuracy of the basis.}
    \label{fig:features}
\end{figure}

\begin{table}
  \begin{tabular}{l l l l l l l l}
   \hline
   Case & $M_\mathrm{tot}[M_\odot]$ & $\eta$ & $\chi_1$ & $\chi_2$ & $\chi_\mathrm{p}$ & $\theta_\mathrm{J}$ & $\alpha_0$\\
   \hline
   (a) & $65.054$ & 0.15  & -0.773 &  0.054 & -0.161 & -0.44  & -0.039\\
   (b) & $62.748$ & 0.144 & -0.772 & -0.153 & -0.134 &  1.084 & 2.773\\
   (c) & $53.375$ & 0.148 & -0.78  &  0.113 & -0.0   &  1.594 & 2.338\\
   (d) & $55.583$ & 0.171 & -0.874 & -0.636 &  0.001 &  1.58  & 1.169\\
   \hline
  \end{tabular}
  \caption{IMRPhenomPv2 parameters for the configurations shown in Fig.~\ref{fig:features}}
  \label{table:features-configurations}
\end{table}

\begin{figure}
\includegraphics[width=0.95\linewidth]{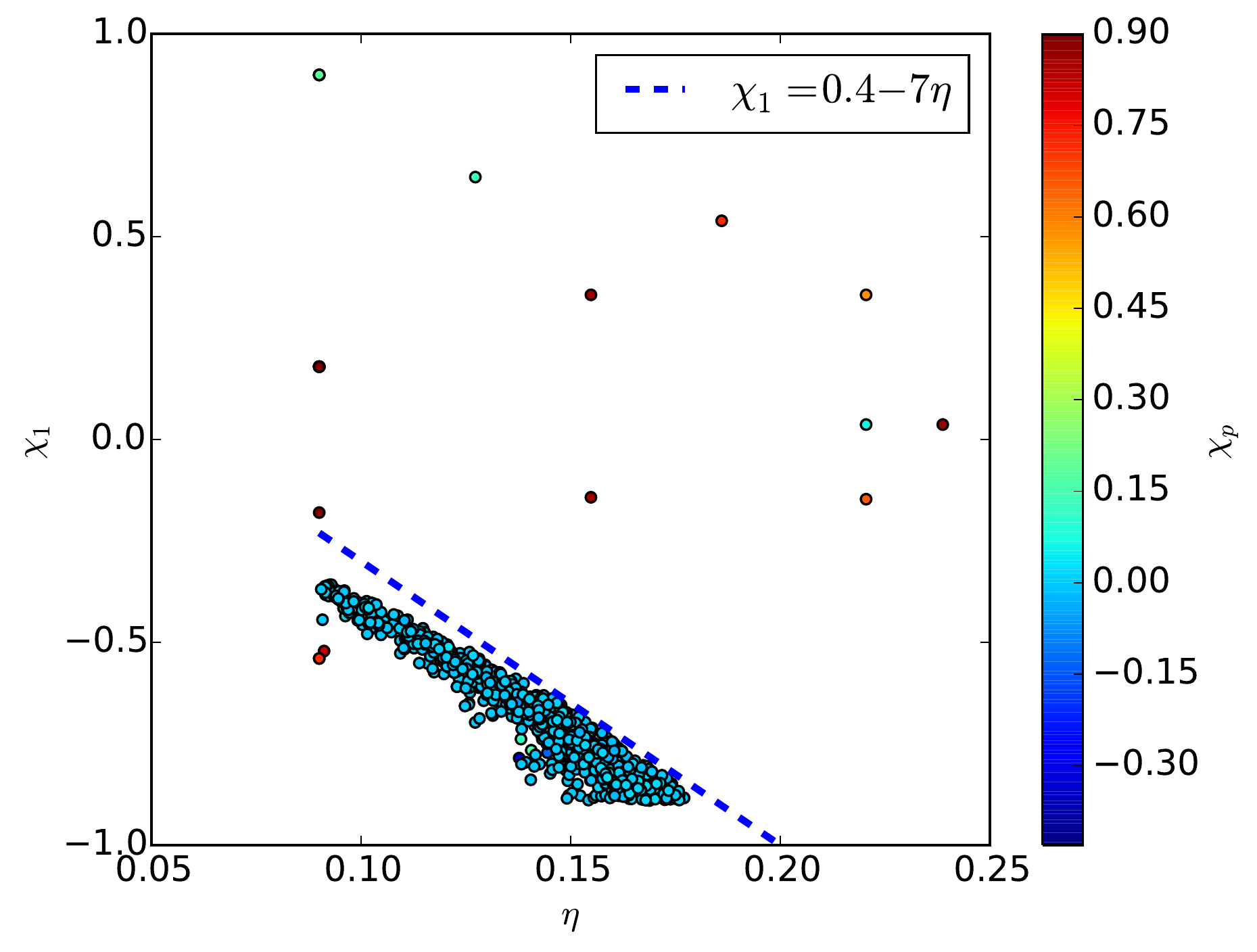}\\
\includegraphics[width=0.95\linewidth]{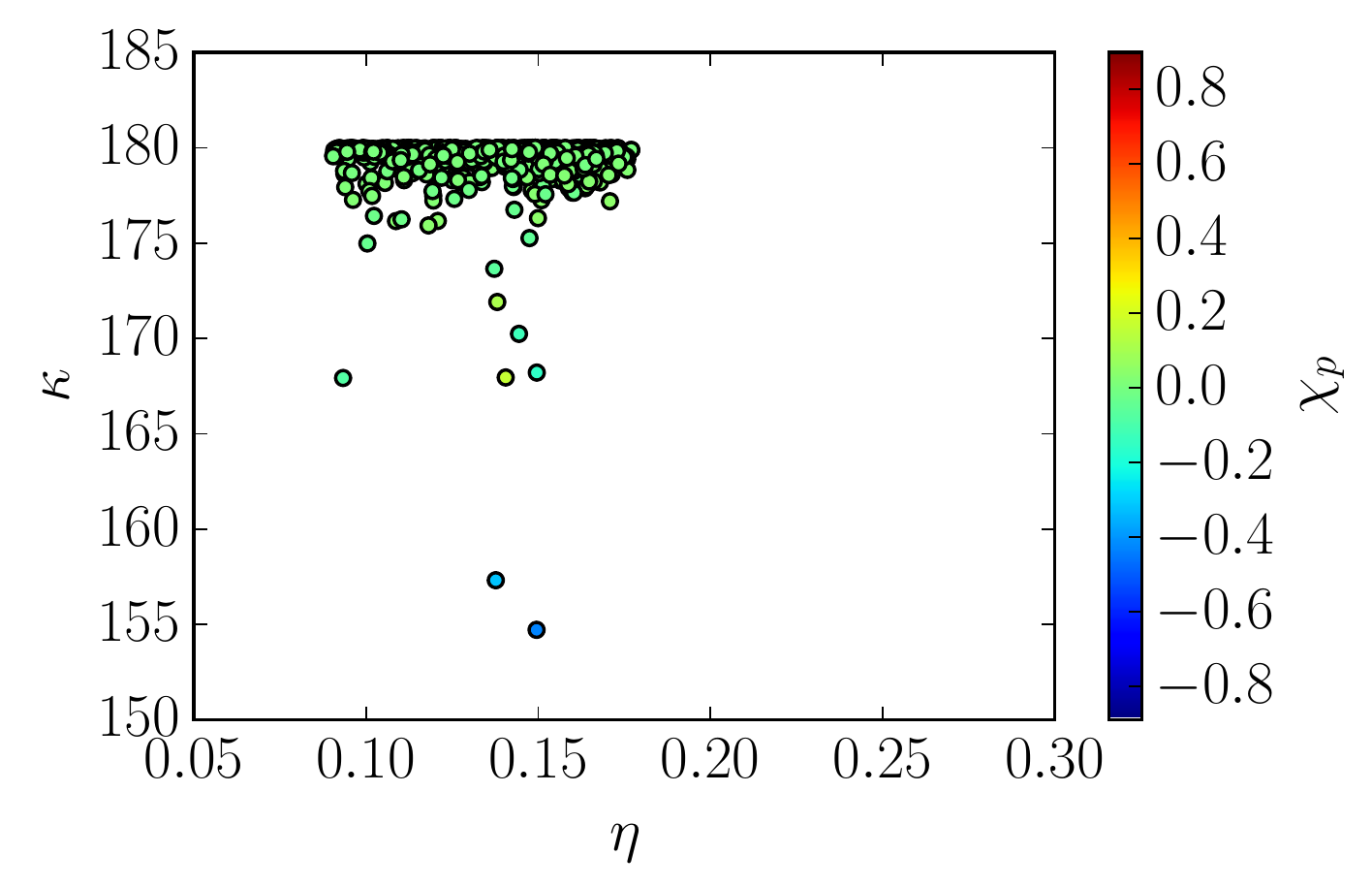}
\caption{{\bf Top}: When applied to the full seven-dimensional parameter space, the greedy algorithm identifies a ``feature cluster" where the model exhibits fast changing (potentially non-smooth) behavior. The cluster directly below the dashed blue line arises for values $\chi_p \approx 0$ large anti-aligned spin $\chi_1$ for unequal mass-ratios. Fig.~\ref{fig:features} shows a few waveforms from this region. {\bf Bottom}: Values of $\kappa$, the angle between the orbital angular momentum L and the total spin $S=S_1 + S_2$, as a function of the symmetric mass ratio $\eta$ for the same cyan ($\chi_p \approx 0$) cluster as shown in the top panel. We observe a clear clustering between $175^\circ$ and $180^\circ$.
}
\label{fig:clusters}
\end{figure}

Below we show an example of the greedy feature detector for case A in Table~\ref{tab:basis_ranges}. Fig.~\ref{fig:clusters} (top) shows a cluster that was identified in the enrichment step of our basis building pipeline. The cluster (cyan circles) corresponds to a subspace that we approximate as $\chi_{1} < 0.4 - 7\eta$. For reasons previously discussed, such clusters are problematic for building ROQs. By removing this cluster from the parameter space in all the cases in Table~\ref{tab:basis_ranges}, we are able to maintain a sparse and accurate basis and empirical interpolant.

The lower panel in Fig.~\ref{fig:clusters} plots the value of $\kappa$, which denotes the angle between $L$ and the total spin $S$ at the reference frequency $f_\mathrm{ref}$, from the $\chi_\mathrm{p} \sim 0$ cluster. We find that the majority of waveforms from this cluster satisfies $175^\circ \leq \kappa \leq 180^\circ$, which is consistent with the condition for the occurrence of transitional precession~\cite{PhysRevD.49.6274} (which, in this case, may or may not be of a physical origin). It was shown in~\cite{PhysRevD.49.6274} that a requirement for the system to undergo transitional precession is $\kappa \geq 164^\circ$. Transitional precession is more likely to occur in binary systems with high mass ratios and initial conditions where the magnitudes of $\vec{L}$ and $\vec{S}$ are similar and point in nearly opposite directions. Such cases are not correctly described by the IMRPhenomPv2 waveform model, and (unphysical) sharp features in this region of the parameter space are identified by the greedy algorithm as shown in the bottom panel of Fig.~\ref{fig:clusters}.

As discussed on Sec.~\ref{sec:phenomp}, the waveform model under consideration, IMRPhenomPv2, does not faithfully model these cases and therefore the occurrence of sharp features in this region of the parameter space may be possible. To illustrate this, Fig.~\ref{fig:features} explicitly shows examples (see Table~\ref{table:features-configurations}) of the abrupt features in the IMRPhenomPv2 amplitudes. For comparison we also plot SEOBNRv3 amplitudes\footnote{The mapping from the general spin information used by SEOBNRv3 to IMRPhenomPv2's internal parameters is surjective. To find parameters for SEOBNRv3 this mapping was inverted with the following choice for the spin components in a frame aligned with ${\hat L}_N$ at $f_\mathrm{ref}=20$Hz: $S_{1x} = \cos(\alpha_0) \chi_\mathrm{p}$, $S_{1y} = \sin(\alpha_0) \chi_\mathrm{p}$, $S_{1z} = \chi_1$, $S_{2x} = S_{2y} = 0$ and $S_{2z} = \chi_2$. Explicitly, the mapping is given by: $(\vec S_1, \vec S_2, {\hat L}_N, f_\mathrm{ref}, m_1, m_2) \twoheadrightarrow (\chi_1, \chi_2, \chi_\mathrm{p}, \theta_\mathrm{J}, \alpha_0, f_\mathrm{ref}, m_1, m_2)$, where ${\hat L}_N = (\sin(\iota), 0, \cos(\iota))$ (in a frame aligned with the view direction), $\iota$ is the angle between ${\hat L}_N$ and the line of sight and $\theta_\mathrm{J}$ is the angle between $\vec J$ and the line of sight.} which behave smoothly for those cases.

\clearpage
\bibliographystyle{apsrev} 
\bibliography{./biblio}

\end{document}

%% file: plots/BinaryFramePrecessing.pdf_tex
\begingroup%
  \makeatletter%
  \providecommand\color[2][]{%
    \errmessage{(Inkscape) Color is used for the text in Inkscape, but the package 'color.sty' is not loaded}%
    \renewcommand\color[2][]{}%
  }%
  \providecommand\transparent[1]{%
    \errmessage{(Inkscape) Transparency is used (non-zero) for the text in Inkscape, but the package 'transparent.sty' is not loaded}%
    \renewcommand\transparent[1]{}%
  }%
  \providecommand\rotatebox[2]{#2}%
  \ifx\svgwidth\undefined%
    \setlength{\unitlength}{369.36455715bp}%
    \ifx\svgscale\undefined%
      \relax%
    \else%
      \setlength{\unitlength}{\unitlength * \real{\svgscale}}%
    \fi%
  \else%
    \setlength{\unitlength}{\svgwidth}%
  \fi%
  \global\let\svgwidth\undefined%
  \global\let\svgscale\undefined%
  \makeatother%
  \begin{picture}(1,0.71421766)%
    \put(0,0){\includegraphics[width=\unitlength]{plots/BinaryFramePrecessing.pdf}}%
    \put(0.05184122,0.00635805){\color[rgb]{0,0,0}\makebox(0,0)[lb]{\smash{$\hat{x}$}}}%
    \put(0.78036181,0.27786186){\color[rgb]{0,0,0}\makebox(0,0)[lb]{\smash{$\hat{y}$}}}%
    \put(0.55048327,0.13207239){\color[rgb]{0,0,0}\makebox(0,0)[lb]{\smash{$m_2$}}}%
    \put(0.03969208,0.4347446){\color[rgb]{0,0,0}\makebox(0,0)[lb]{\smash{$m_1$}}}%
    \put(0.23555309,0.68459753){\color[rgb]{0,0,0}\makebox(0,0)[lb]{\smash{$\hat{J} \equiv \hat{z}$}}}%
    \put(0.46691439,0.54493292){\color[rgb]{0,0,0}\makebox(0,0)[lb]{\smash{$\vec{L}$}}}%
    \put(0.34289234,0.18309871){\color[rgb]{0,0,0}\makebox(0,0)[lb]{\smash{$\alpha_0$}}}%
    \put(0.37152233,0.49295189){\color[rgb]{0,0,0}\makebox(0,0)[lb]{\smash{$\iota$}}}%
    \put(0.24654549,0.38181653){\color[rgb]{0,0,0}\makebox(0,0)[lb]{\smash{$\theta_J$}}}%
    \put(0.03821708,0.26867035){\color[rgb]{0,0,0}\makebox(0,0)[lb]{\smash{$\hat{N}$}}}%
    \put(0.10631145,0.53034058){\color[rgb]{0,0,0}\makebox(0,0)[lb]{\smash{$\chi_1$}}}%
    \put(0.59538613,0.20219454){\color[rgb]{0,0,0}\makebox(0,0)[lb]{\smash{$\chi_2$}}}%
    \put(0.1714628,0.42642703){\color[rgb]{0,0,0}\makebox(0,0)[lb]{\smash{$\chi_p$}}}%
  \end{picture}%
\endgroup%